\documentclass[12pi,epsfig,elsart]{article}
\usepackage{epsfig}

\newcommand{\be}{\begin{equation}}
\newcommand{\ee}{\end{equation}}
\newcommand{\bea}{\begin{eqnarray}}
\newcommand{\eea}{\end{eqnarray}}
\newcommand{\bean}{\begin{eqnarray*}}
\newcommand{\eean}{\end{eqnarray*}}

\begin{document}
\begin{center}
{\Large \bf Resonances Formed by $\bar pp$ and Decaying into
	$\pi^0\pi^0\eta$ for Masses 1960 to 2410 MeV }
\vskip 5mm
{A.V. Anisovich$^c$, C.A. Baker$^a$, C.J. Batty$^a$, D.V. Bugg$^b$,  C. Hodd$^b$,
J. Kisiel$^d$, V.A. Nikonov$^c$, A.V. Sarantsev$^c$, V.V. Sarantsev$^c$, 
I. Scott$^b$, B.S.~Zou$^{b}$ \footnote{Now at IHEP, Beijing 100039, China} }\\
{\normalsize $^a$ \it Rutherford Appleton Laboratory, Chilton, Didcot OX11 0QX,UK}\\
{\normalsize $^b$ \it Queen Mary and Westfield College, London E1\,4NS, UK}\\
{\normalsize $^c$ \it PNPI, Gatchina, St. Petersburg district, 188350, Russia}\\
{\normalsize $^d$ \it University of Silesia, Katowice, Poland}
\\ [3mm]
\end {center}

\begin{abstract}

Data on $\bar pp$ annihilation in flight into $\pi^0\pi^0\eta$ are
presented for nine beam momenta 600
to 1940 MeV/c. The strongest four intermediate states are found to be
$f_2(1270)\eta$, $a_2(1320)\pi$, $\sigma\eta$ and
$a_0(980)\pi$.
Partial wave  analysis is performed mainly to
look for resonances formed by $\bar pp$ and decaying into $\pi^0\pi^0\eta$
through these intermediate states.
There is evidence for the following $s$-channel $I = 0$ resonances :
two $4^{++}$ resonances with mass and width
(M, $\Gamma$) at ($2044$, $208$) MeV and
($2320\pm 30$, $220\pm 30$) MeV; three $2^{++}$ resonances at
($2020\pm 50$, $200\pm 70$) MeV, ($2240\pm 40$, $170\pm 50$) MeV and
($2370\pm 50$, $320\pm 50$) MeV; two $3^{++}$ resonances at
($2000\pm 40$, $250\pm 40$) MeV and ($2280\pm 30$, $210\pm 30$) MeV;
a $1^{++}$ resonance at
($2340\pm 40$, $340\pm 40$) MeV;
and two $2^{-+}$ resonances at
($2040\pm 40$, $190\pm 40$) MeV and ($2300\pm 40$, $270\pm 40$) MeV.

\end{abstract}


\section{Introduction}

The Crystal Barrel detector is being used to make a systematic study of the mass
region 1960 to 2410 MeV in $\bar pp$ annihilation in flight at LEAR, with
$\bar p$ beams of momenta 600 to 1940 MeV/c.
The objective is to study resonances in the formation process, i.e. the
$s$-channel.
Here we study data in $\bar pp \to \pi ^0 \pi ^0 \eta$ for resonances
decaying to $a_2(1320)\pi ^0$, $f_2(1270)\eta$, $f_0(1500)\eta$,
$f_0(980)\eta$, $a_0(980)\pi ^0$ and $\sigma \eta$.
We use $\sigma$ to denote the broad $\pi \pi $ S-wave amplitude up to $\sim
1860$ MeV.
The present results have been presented briefly in the form of a letter [1] and
here we give full details of the experimental techniques and analysis.
Further studies of $\eta \eta \pi ^0$ have been presented elsewhere [2,3], and
work is in progress on other channels such as $3\pi ^0$ and $\pi ^0 \pi ^0 \eta
'$.

From earlier work, it is known that the mass range we explore
contains many resonances [4];
a detailed study of $\bar pp \to \pi ^- \pi ^+$ using a polarised target has
provided much of the current evidence [5,6].
The $f_4(2050)$ is well known, and from the quark model of meson resonances
one expects that it will be
accompanied by $f_3$ and $f_2$ resonances close-by in mass.
We shall indeed provide evidence for these resonances and a further one with
quantum numbers $J^{PC} = 2^{-+}$ and similar mass.
At higher masses, towards the top of the LEAR range, there has been
evidence for $f_4(2300)$ and $f_2(2340)$ [4],
and it is anticipated from the Veneziano model
[7] that there is likely to be a tower of resonances
around this mass.
We shall provide evidence for such
states with quantum numbers $4^+$, $3^+$, $2^+$, $1^+$ and $2^-$.

These resonances are anticipated $\bar qq$ states.
This mass range is also likely to contain glueballs with quantum numbers
$0^{-+}$ and $2^{++}$, predicted in the mass range 2000--2400 MeV by
various theoretical models [8,9,10].
Hybrids may also be present.
Decays of these exotic resonances to $\eta$ and $\sigma$ seem to be favoured in
$f_0(1500)$ decay [11], charmonium decay and $J/\Psi$ radiative decays [12].
Hence the $\eta f_2(1270)$ and $\eta \sigma$ channels are of particular
interest.

The layout of this paper will be as follows.
In section 2, the procedure for data processing and event selection is
outlined; the data are presented and their gross features are discussed.
Section 3 gives the formalism used for the partial wave analysis.
Section 4 gives the results for partial wave amplitudes.
Then, in Section 5 we fit partial waves to resonances.
Finally, Section 6 provides a summary.

\section {Experiment and Data Processing}
The data were taken at LEAR by the Crystal Barrel Collaboration, using a
trigger
on neutral final states at nine beam momenta from 600 to 1940 MeV/c.
An average of $9\times 10^6$ triggers were taken
at each momentum. The detector has been described fully in an earlier
publication [13].

A liquid hydrogen target 4.4 cm long is surrounded at increasing radii by a
silicon vertex detector, a multiwire chamber for triggering, a jet drift
chamber to detect charged particles and finally 1380
CsI crystals to detect photons.
The present data were taken with a trigger demanding a neutral final state.
For this purpose, the silicon vertex detector, multiwire chamber and
jet drift chamber were used simply to veto charged particles.

The barrel of CsI crystals covers 98\% of $4\pi$ solid angle.
Crystals are 16 radiation lengths long and point towards the target.
The angular resolution is $\sim \pm 20$ mrad in both polar angle and azimuth.
The detection efficiency is high for photons down to energies below 20 MeV.
The energy resolution $\Delta E$ is given by $\Delta E/E = 0.025/E^{1/4}$,
where $E$ is in GeV.

The incident $\bar p$ beam was pure and monoenergetic with momentum spread
$\Delta p /p < 0.1\%$.
Incident antiprotons were defined by a coincidence between a small
proportional counter P and a 5mm diameter silicon counter, Si.
Two veto counters, 20 cm downstream of the hydrogen target, were used to provide
a first level trigger $P.Si.\bar V$ identifying interactions in the target.
The beam intensity was typically $2 \times 10^5$ $\bar p$/s and at times was
twice this.
The interaction rate in the target (excluding $\bar pp$ elastic scattering,
where the forward $\bar p$ generally counted in the veto counter) was typically
3KHz. Of this, $\sim 1-2\%$ consisted of neutral final states,
so the trigger rate
for all-neutral events was 20--60 Hz.
In order to filter out events which obviously fail to conserve energy,
the total energy in the CsI crystals was summed on-line
[14]; a fast trigger rejected those events with  total energy falling
$\sim 200$ MeV or more below that of  $\bar pp$ annihilation.

The absolute normalisation is derived from beam counts $P.Si$,
target length and density, the number of
detected events and a Monte Carlo simulation of reconstruction efficiency in
the CsI barrel.
Details of this normalisation are given in a paper on the $\pi ^0 \pi ^0$
final state [15].
A dependence of the reconstruction efficiency on beam rate is observed, and
the normalisation has to be obtained from an extrapolation to zero beam rate.
The normalisation uncertainty is estimated as $\pm 3\%$ from 1800 to 1050
MeV/c and increases to $\pm 6\%$ at 900 and 600 MeV/c.
Data at 1940 MeV/c were taken in separate, earlier runs, and have an estimated
uncertainty of $\pm 10\%$ in normalisation.
There is in addition an overall normalisation uncertainty of
$\pm 2.4\%$ from the target length, common to all momenta.

\subsection {Data Selection}
A large number of alternative prescriptions have been examined for selecting
events.
At high momenta, one of the problems is that photons from $\pi ^0$ decay
sometimes merge into a single shower.
Conversely, one shower sometimes splits into a primary shower and a nearby
secondary shower, caused by Compton scattering.
The probability that this occurs is $\sim 10 \%$ per photon.
In early studies, an attempt was made to salvage $\eta \pi ^0 \pi ^0$ events
from $5\gamma$ or $7\gamma$ final states.
However, the gain in statistics was small ($\sim 15\%)$ and the penalty was an
increase in backgrounds. Eventually, it was decided to retain only events
containing exactly 6 photon showers.

\begin{table}[htp]
\begin{center}
\begin{tabular}{cccc}
\hline
Beam Momentum & Number of Events &  Reconstruction & Cross Section  \\
(MeV/c) & & Efficiency (\%) & ($\mu b$)\\\hline
 600 &  20385 & 26.3 & $71.9\pm 3.6$ \\
 900 & 112476 & 25.4 & $83.2\pm 4.9$ \\
1050 &  86238 & 24.9 & $78.9\pm 2.3$ \\
1200 & 124581 & 24.2 & $68.6\pm 3.0$ \\
1350 &  81454 & 23.4 & $54.4\pm 2.3$ \\
1525 &  57714 & 22.7 & $56.5\pm 1.8$ \\
1642 &  65984 & 21.9 & $53.2\pm 2.5$ \\
1800 &  71738 & 20.8 & $43.8\pm 1.5$ \\
1940 &  75325 & 19.8 & $37.0\pm 3.7$ \\\hline
\end {tabular}
\caption { Numbers of selected events, reconstruction efficiency and cross
sections for $\bar pp\to\pi^0\pi^0\eta$ with $\eta\to\gamma\gamma$.}
\end{center}
\end{table}

Data are fitted kinematically to a large number of physics channels: 43 for
$(4-8)\gamma $.
In order to assess branching ratios to every channel and cross-talk between
them, we generate at least 20,000 Monte Carlo events for every one of the 43
fitted channels, using GEANT.
In the first approximation, events fitting the correct channel determine
the reconstruction efficiency $\epsilon _i$ in each channel.
Events fitting the wrong channel estimate the probability of cross-talk
$x_{ij}$ between channels $i$ and $j$.
More exactly, we solve
a set of 43 x 43 simultaneous equations containing on the left-hand
side the observed number of fitted data events $D_i$, and on the right-hand
side reconstruction efficiencies and true numbers of created  events $N_i$
in every channel and
terms allowing for cross-talk $x_{ij}$ between channels:
\begin {equation}
D_i = \epsilon _iN_i + \sum _{j\ne i}~x_{ij}N_j.
\end {equation}
The solution is constrained so that the numbers of real events, $N_i$, in every
channel are positive or zero.

\vskip -10mm
\begin{figure}[htbp]
\begin{center}
\epsfysize=7cm
\epsffile{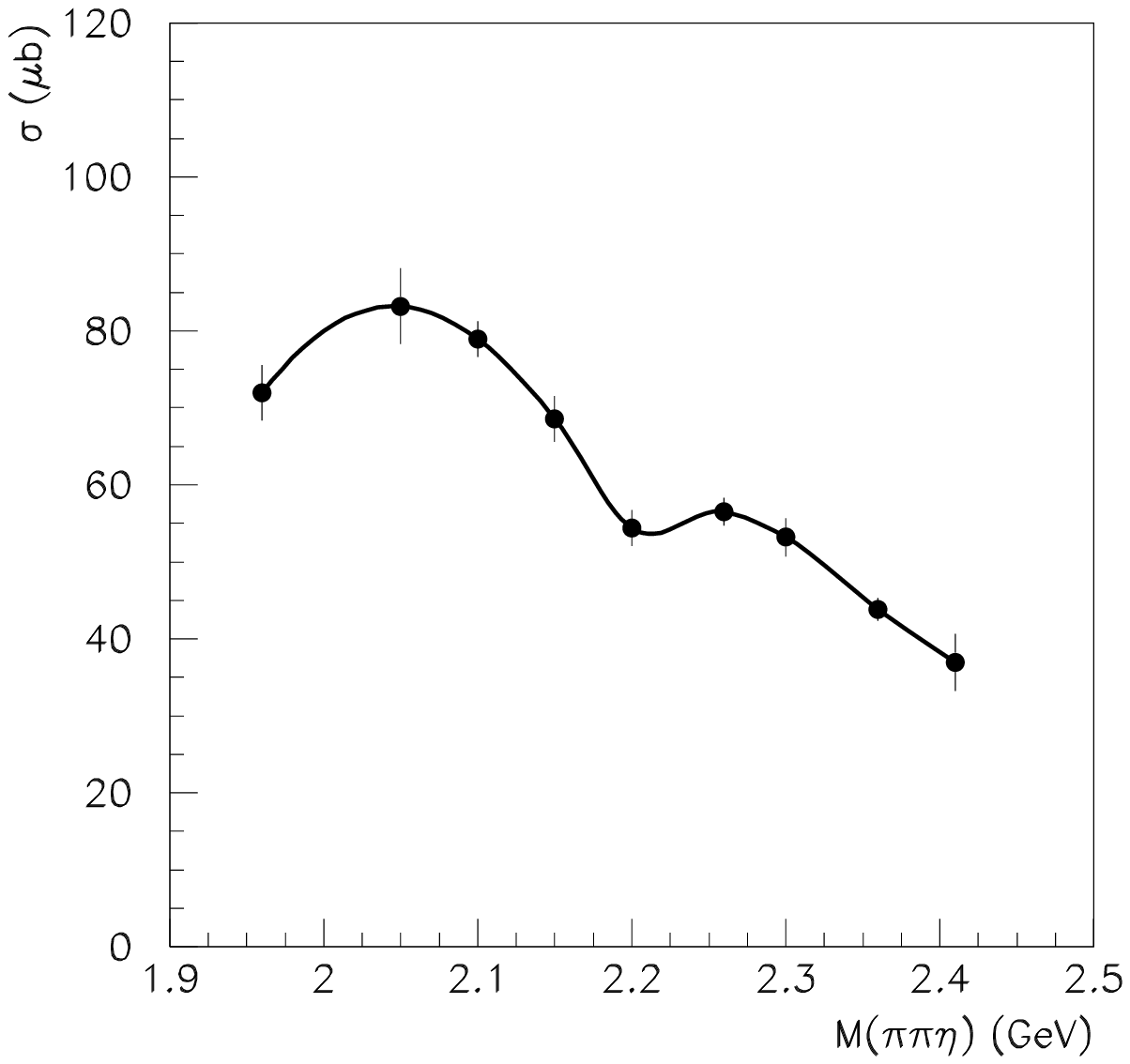}
\vskip -70.45mm
\epsfysize=7cm
\epsffile{XS.PS}
\vskip -8mm
\caption{Cross section for $\bar pp\to\pi^0\pi^0\eta$ with
$\eta\to\gamma\gamma$. }
\end{center}
\end{figure}
This procedure is carried out for a variety of confidence levels (1, 5, 10,
20\%) and using a wide variety of selection procedures.
A choice is then made, optimising the ratio of signal to background.
We find that this ratio is not very sensitive to confidence level over the
range 5--20\% for $\eta \pi ^0 \pi ^0$ events.

Among $6-8$ photon events, the four largest channels are $4\pi^0$,
$3\pi^0$, $\pi^0\pi^0\eta$ and $\pi^0\pi^0\omega$ with $\omega\to\pi^0\gamma$.
The relative branching
ratios for these channels are roughly 1.1 : 1 : 0.4 : 0.4 at 1800 MeV/c.
To select the $\pi^0\pi^0\eta$ channel, we demand exactly 6 photons
satisfying a 7C kinematic fit with confidence level $>10 \%$; events fitting
$3\pi ^0$  with confidence level $>0.01\%$ are rejected,
and also those few events fitting $\pi^0\pi^0\eta'$, $\pi^0\eta\eta$,
$\pi^0\eta\eta'$ and $3\eta$ with confidence level larger than that for
$\pi^0\pi^0\eta$. The  Monte Carlo simulation shows that
the worst backgrounds arise from $\omega\pi^0\pi^0$, ($\omega\to\pi^0\gamma$)
when one photon is lost, and from
$4\pi^0$ events when two photons are lost.
Residual  backgrounds from these two processes are 1.5\% and 0.8\%
respectively at 1800 MeV/c.
Including other small backgrounds, the total is  $3.0\pm 0.3\%$ at 1800 MeV/c.
For lower beam momenta, the background increases slightly. At 600 MeV,
the total is $4.0\pm 0.4$ with the worst backgrounds from $\omega\pi^0\pi^0$
(1.7\%), $4\pi^0$ (0.9\%) and $\omega\omega$ (0.9\%).
Table 1 summarises numbers of selected events, the reconstruction efficiency
and cross sections. Statistics at 600 MeV/c are lower because most
data were taken without the threshold cut on total energy in the trigger.
The cross sections for the $\eta \pi ^0 \pi ^0$ channel are also shown in Fig.1.
There are clear enhancements at low mass and around 2200--2300 MeV.
Note that for a constant amplitude
the cross sections should decrease steadily as the energy increases, see equn
(23) below.

\begin{figure}
\begin{center}
\epsfysize=9cm
\epsffile{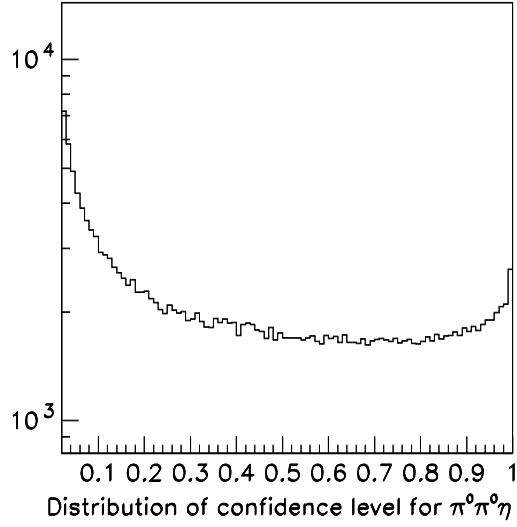}
\vskip -90.45mm
\epsfysize=9cm
\epsffile{CL1200.PS}
\end{center}
\vskip -8mm
\caption{Distribution of confidence level for $\bar
pp\to\pi^0\pi^0\eta$ events at beam momentum 1.2 GeV/c. }
\end{figure}

Fig. 2 shows the confidence level (CL) distribution for data of beam
momentum at 1.2 GeV/c.
The slight peak at high confidence level arises from events where all particles
emerge close to the beam direction, with the result that the vertex is
poorly defined.
We apply no cut on the coordinate of the vertex along the beam direction,
so as to avoid biasing the data selection.
The rise at low confidence levels is followed accurately down to 10\%
by the Monte Carlo simulation;
it arises from overlapping showers in the CsI detectors.

In order to illustrate the cleanliness of the $\eta$ signal, we have made an
additonal  fit to $\pi ^0 \pi ^0 \gamma \gamma$.
Fig. 3 then shows the mass distribution of $\gamma \gamma$ pairs in the
vicinity of the $\eta$ peak for $CL(\pi^0\pi^0\gamma\gamma)>0.1$ with
$CL(\pi^0\pi^0\eta)>0.0001$ at beam momentum 1.2 GeV/c.
The $\eta$ peak is well centred at the correct mass, 547.5 MeV and the
background under the $\eta$ signal is compatible with that expected from the
Monte Carlo simulation.

\begin{figure}[htbp]
\begin{center}
\epsfysize=9cm
\epsffile{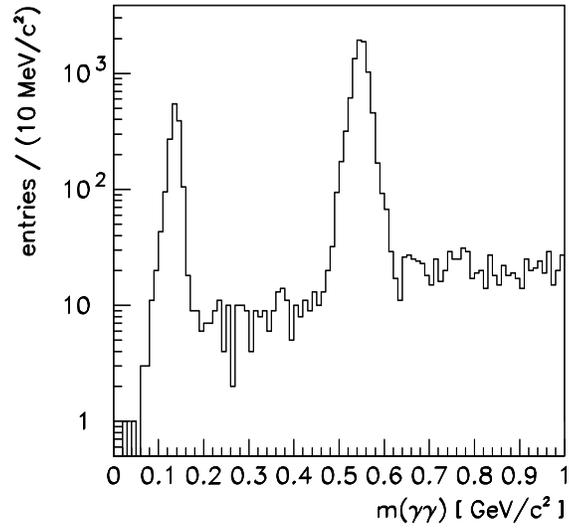}
\vskip -90.45mm
\epsfysize=9cm
\epsffile{GG1200.PS}
\end{center}
\vskip -8mm
\caption{Mass distribution of $\gamma\gamma$ pairs for
$CL(\pi^0\pi^0\gamma\gamma)>0.1$ with
$CL(\pi^0\pi^0\eta)>0.0001$ at beam momentum 1.2 GeV/c.}
\end{figure}

\begin{figure}[htbp]
\begin{center}
\epsfysize=18cm
\epsffile{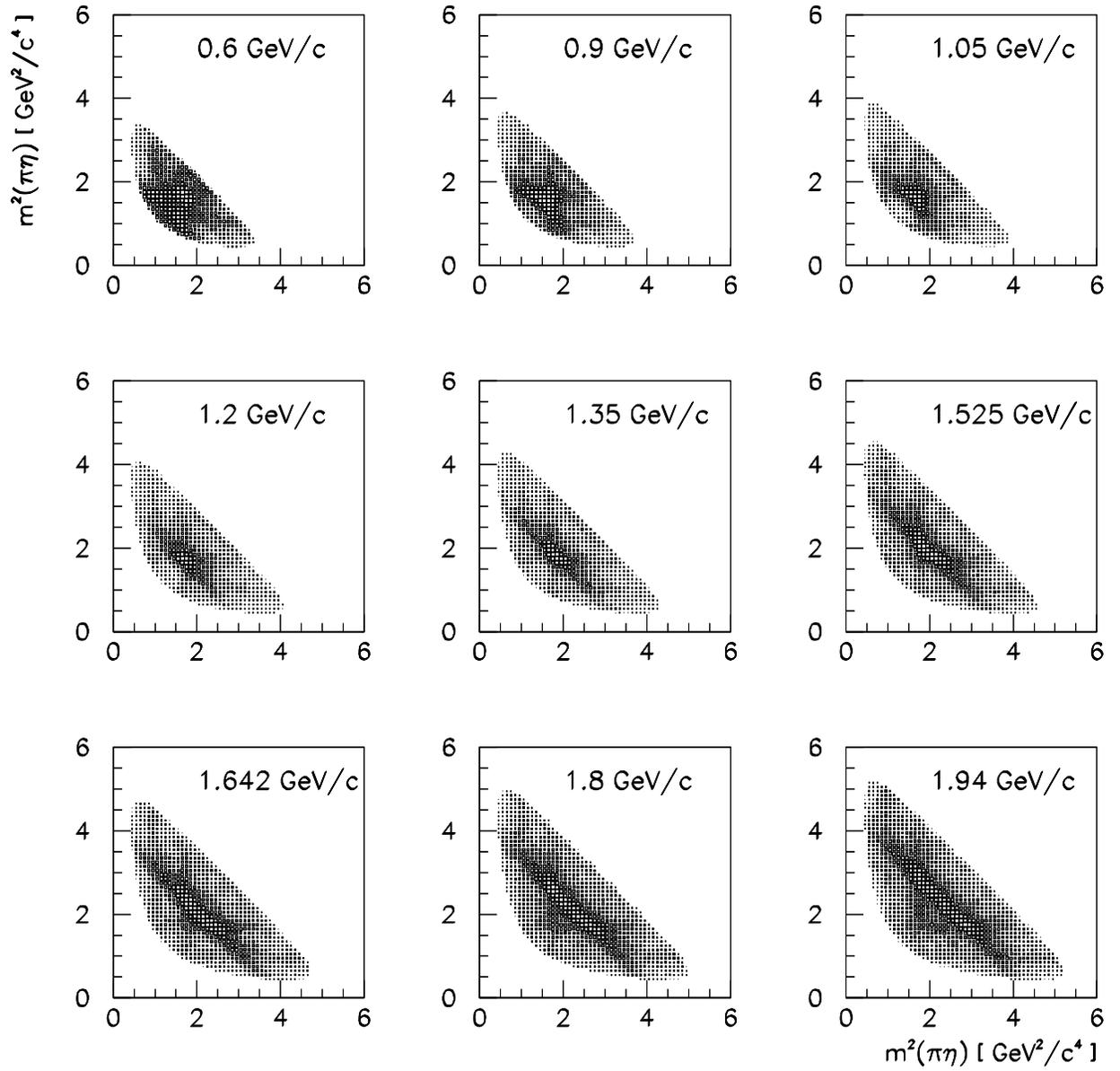}
\vskip -180.45mm
\epsfysize=18cm
\epsffile{DALITZ.PS}
\vskip -8mm
\end{center}
\caption{Dalitz plots for $\bar pp\to\pi^0\pi^0\eta$ at incident beam
momenta $0.6 - 1.94$ GeV/c.}
\end{figure}

\subsection {Features of the Data}
Fig. 4 shows Dalitz plots at the nine available momenta and Figs. 5 and 6
projections on to $\eta \pi$ mass and $\pi \pi $ mass.
The most prominent feature of the Dalitz plot consists of a diagonal
band due to $f_2(1270)\pi ^0$.
There are weaker horizontal and vertical bands due to $a_0(980)\pi$ and
$a_2(1320)\pi$.
The $f_2(1270)\eta$ signal grows with respect to $a_2(1320)\pi$ as the beam
momentum rises;
this is a natural consequence of the increasing phase space for
$f_2(1270)\eta$, whose threshold is at 1820 MeV.
Very weak peaks are visible in the $\pi \pi$ mass
projection of Fig. 6 due to $f_0(1500)\eta$ and $f_0(980)\eta$.
In addition, there is some slowly varying contribution
covering the whole Dalitz plots; it may come from the broad $\sigma$, i.e.,
$f_0(400-1200)$ in the Particle Data Tables [4].
We adjust fitted masses and widths of $f_2(1270)$ and $a_2(1320)$ by a
few MeV from PDG values in order to achieve the optimum fits.
This is because our main aim is to fit the production and decay angular
distributions of these resonances.

\vskip -8mm
\begin{figure}[htbp]
\begin{center}
\epsfysize=11cm
\epsffile{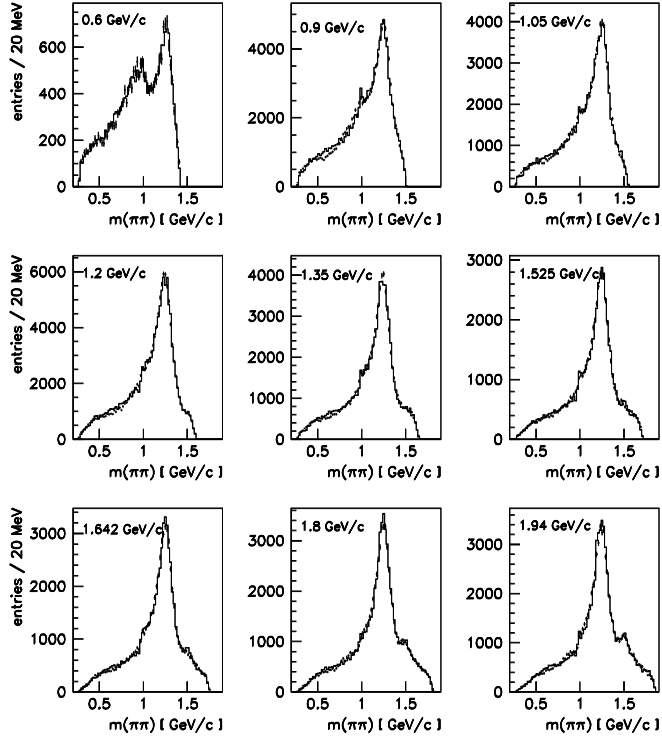}
\vskip -110.45mm
\epsfysize=11cm
\epsffile{PROJ1.PS}
\vskip -8mm
\end{center}
\caption{Data and fit (histogram) of invariant mass spectra for $\pi^0\pi^0$
(1 entry/event). }
\end{figure}

\vskip -2mm
\begin{figure}[htbp]
\begin{center}
\epsfysize=10.8cm
\epsffile{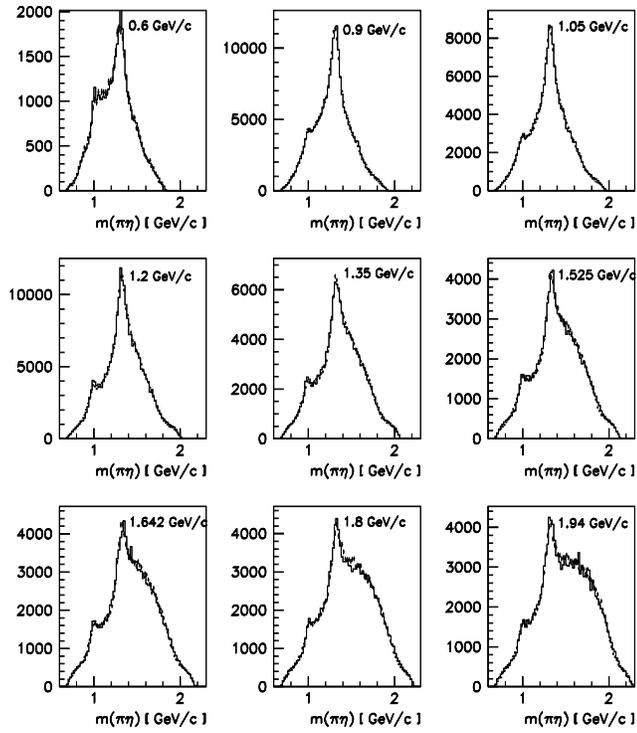}
\vskip -108.45mm
\epsfysize=10.8cm
\epsffile{PROJ2.PS}
\vskip -8mm
\end{center}
\caption{Data and fit (histogram) of invariant mass spectra for
$\pi^0\eta$ (2 entries/event). }
\end{figure}

Figs. 7 and 8 show differences on the Dalitz plot between fit and data.
There are small systematic discrepancies at the extreme right-hand edge of the
Dalitz plot near an $\eta \pi$ mass of 1450 MeV.
This discrepancy may be due to $a_0(1450)$ or $a_2(1660)$ or $\hat \rho
(1405)$.
The effect is small and cannot be analysed unambiguously into partial waves.
Fits including these components have almost no effect on the main components of
the fit, with the exception of $\eta \sigma$, which covers the whole Dalitz
plot and can absorb other small, ill-defined contributions.

\begin{figure}[htbp]
\begin{center}
\epsfysize=11cm
\epsffile{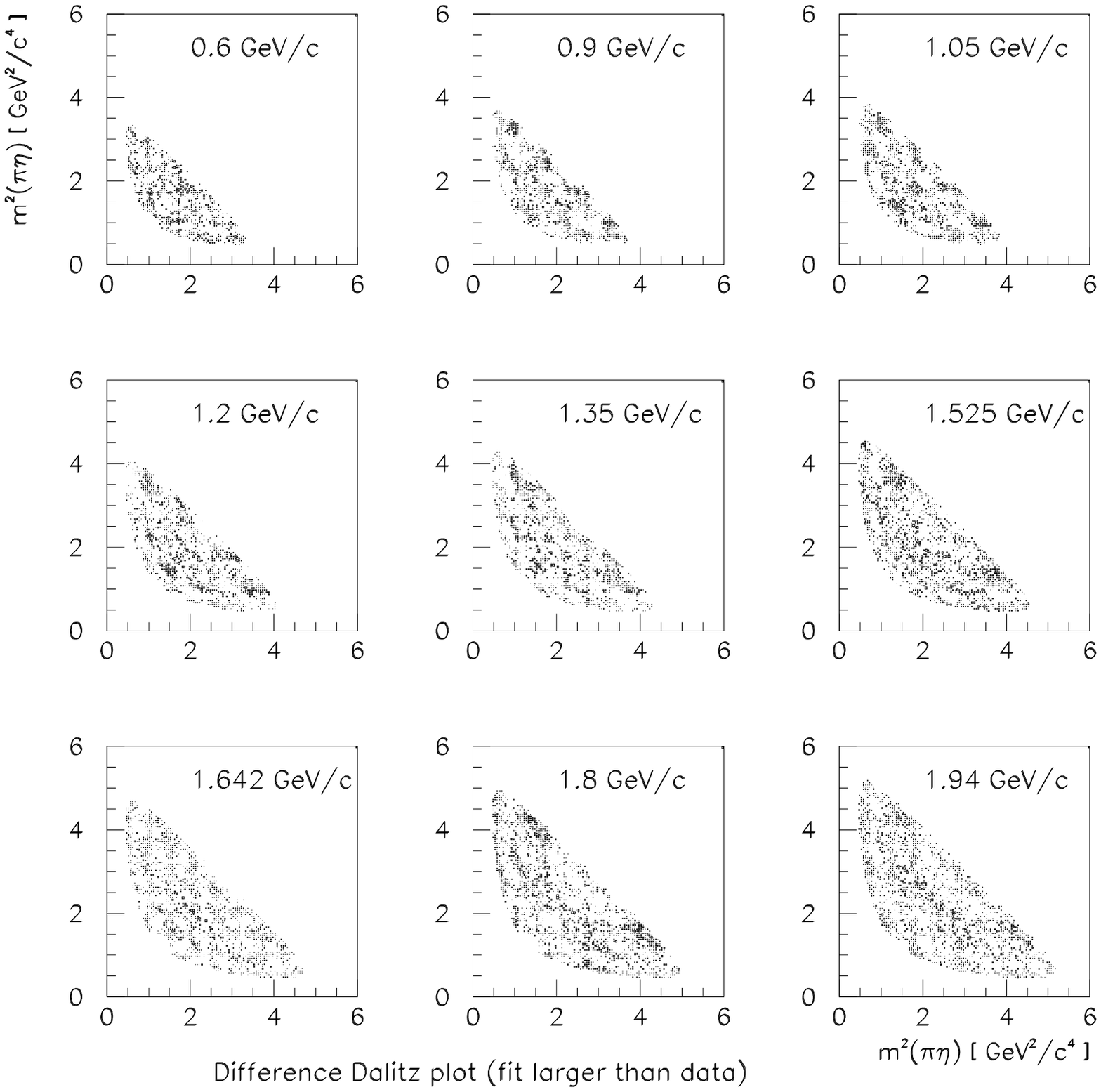}
\vskip -110.45mm
\epsfysize=11cm
\epsffile{DIFF1.PS}
\end{center}
\vskip -8mm
\caption{Difference between Dalitz plots of fit and data where fit $ > $ data.}
\end{figure}

\begin{figure}[htbp]
\begin{center}
\epsfysize=11cm
\epsffile{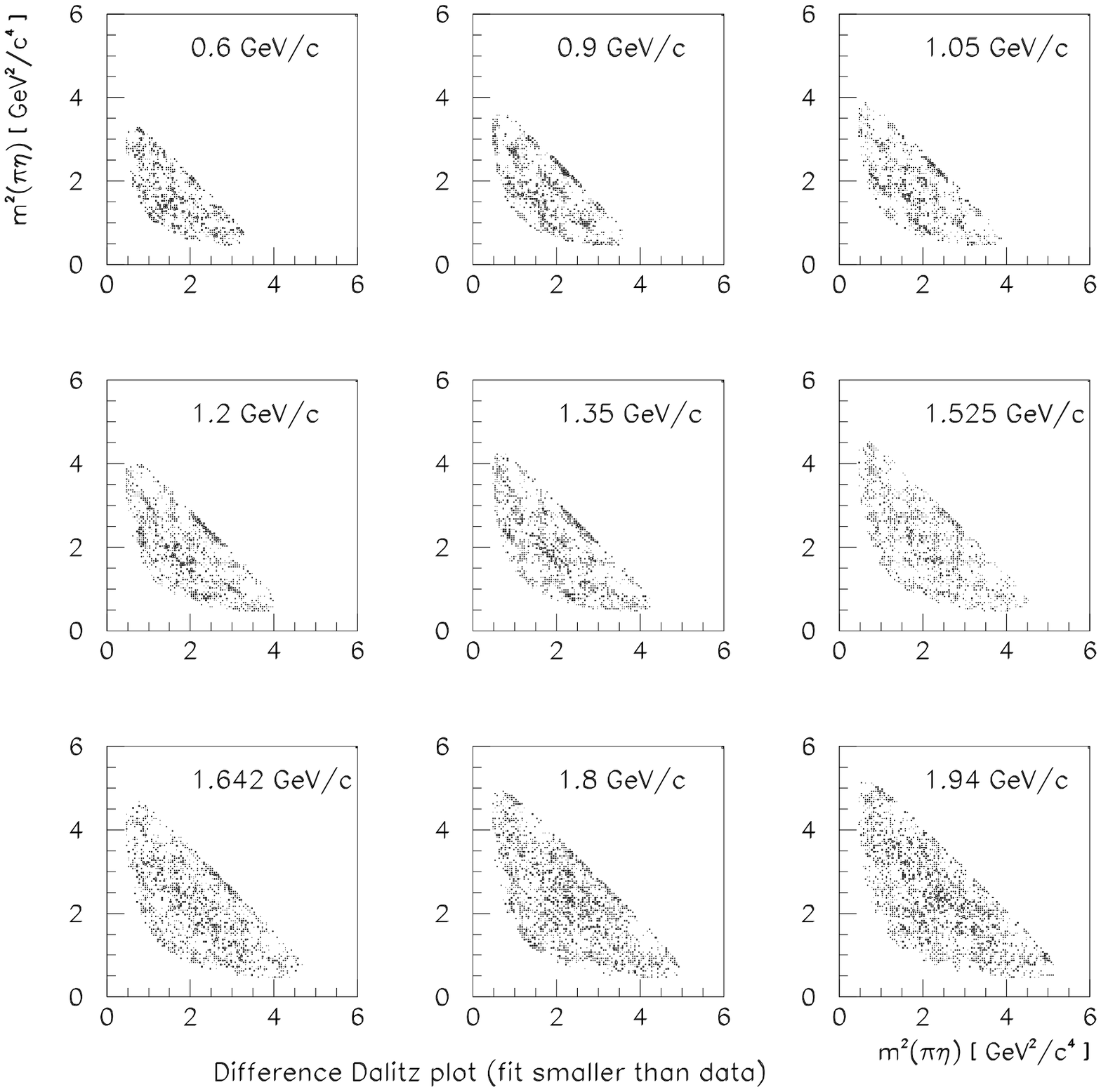}
\vskip -110.45mm
\epsfysize=11cm
\epsffile{DIFF2.PS}
\end{center}
\vskip -8mm
\caption{Difference between Dalitz plots of fit and data where fit $ < $ data.}
\end{figure}

Figs. 9 and 10 show production angular distributions (after acceptance
correction) for events lying in the $f_2(1270)$ mass band ($1275\pm 100$ MeV)
and for events lying in the $a_2(1320)$
mass band ($1320\pm 50$ MeV).
It is immediately obvious that high orbital angular momenta are involved for
both $f_2\eta$ and $a_2\pi$ at the higher beam momenta. The histograms show
results of the partial wave fit described below.
\begin{figure}[htbp]
\begin{center}\hspace*{-0.cm}
\epsfysize=14cm
\epsffile{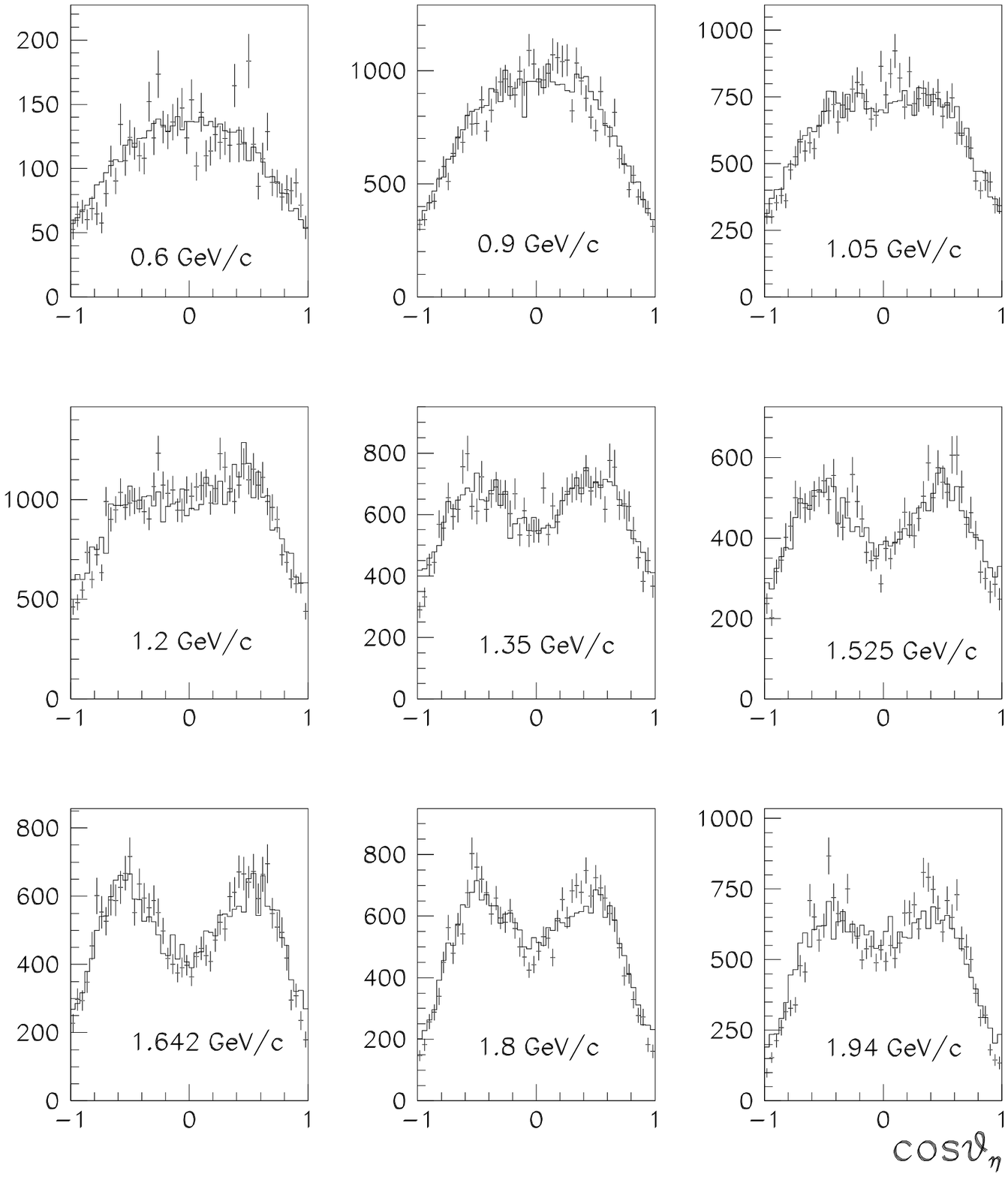}
\vskip -140.45mm
\hskip 1.35mm
\epsfysize=14cm
\epsffile{ANGLE1.PS}
\end{center}
\caption{Data and fit (histogram) of angular distribution
$d\sigma/d~cos\theta_\eta$
for $M_{\pi\pi}$ between 1175 and 1375 MeV  (1 entry/event). }
\end{figure}

\begin{figure}[htbp]
\begin{center}\hspace*{-0.cm}
\epsfysize=14cm
\epsffile{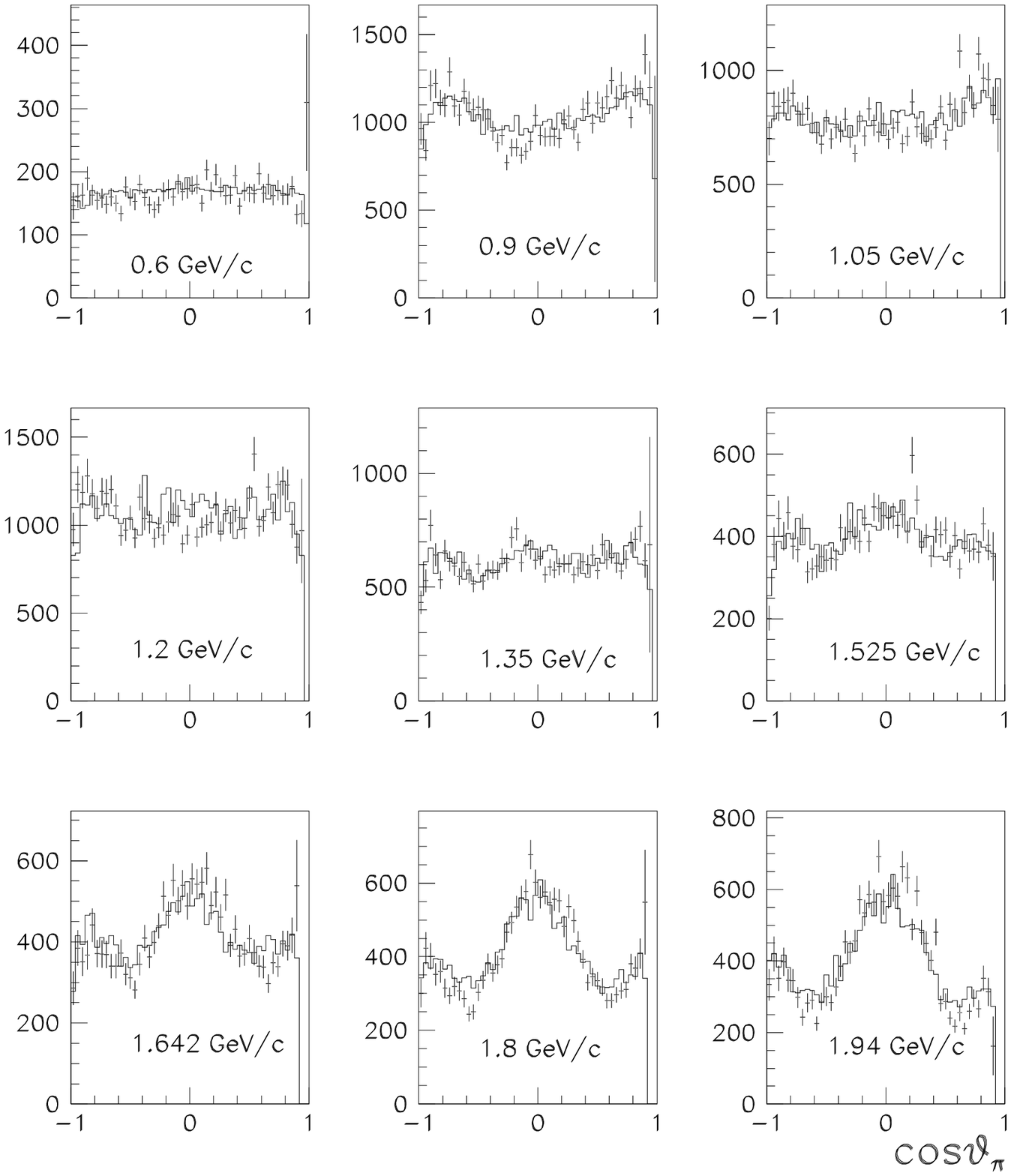}
\vskip -140.45mm
\hskip 1.35mm
\epsfysize=14cm
\epsffile{ANGLE2.PS}
\end{center}
\vskip -6mm
\caption{Data and fit (histogram) of angular distribution
$d\sigma/d~cos\theta_\pi$
for $M_{\eta\pi}$ between 1270 and 1370 MeV  (2 entries/event). }
\end{figure}

\section{Formalism for Partial Wave Analysis}

For the $\pi^0\pi^0\eta$ final state, possible $\bar pp$ initial singlet
states are $0^{-+}$, $2^{-+}$, $4^{-+}$ etc;
for $\bar pp$ spin triplet, allowed  states are
$1^{++}$, $2^{++}$, $3^{++}$, $4^{++}$, $5^{++}$ etc.
For our case with center-of-mass energies below 2.41 GeV,
only $0^{-+}$, $2^{-+}$, $1^{++}$, $2^{++}$, $3^{++}$ and $4^{++}$ are
expected to be significant [6] and this has been confirmed in our analysis;
$4^{-+}$ has been tried, but is not significant.
The corresponding $\bar pp$
states with total angular momentum J, orbital angular momentum L and total
spin angular momentum S in the usual contracted form $^{2S+1}L_J$ are:
$^1S_0$ for $0^{-+}$, $^1D_2$ for $2^{-+}$, $^3P_1$ for $1^{++}$,
$^3P_2$ or $^3F_2$ for $2^{++}$, $^3F_3$ for $3^{++}$, and $^3F_4$ or $^3H_4$
for $4^{++}$.

Let us choose the reaction rest frame with the z axis along the $\bar p$
beam direction. Then the squared modulus of the total transition amplitude
is the following [16]:
\bea
I&=&|A_{0^{-+}}+A_{2^{-+}}|^2+|A^{M=1}_{1^{++}}+A^{M=1}_{3^{++}}|^2
+|A^{M=-1}_{1^{++}}+A^{M=-1}_{3^{++}}|^2 \nonumber\\
& & +|A^{M=0}_{2^{++}}+A^{M=0}_{4^{++}}|^2
+|A^{M=1}_{2^{++}}+A^{M=1}_{4^{++}}|^2+|A^{M=-1}_{2^{++}}+A^{M=-1}_{4^{++}}|^2
\\\nonumber
& & +2Re[(A^{M=1}_{2^{++}}+A^{M=1}_{4^{++}})
(A^{M=1}_{1^{++}}+A^{M=1}_{3^{++}})^*
-(A^{M=-1}_{2^{++}}+A^{M=-1}_{4^{++}})(A^{M=-1}_{1^{++}}+A^{M=-1}_{3^{++}})^*]
\eea
where M is the spin projection on the z-axis in the initial state.
The absence of M=0 for $1^{++}$
and $3^{++}$ is due to the vanishing of the Clebsch-Gordon (CG) coefficient
$(J=2n+1,M_J=0|L=2n+1,M_L=0; S=1, M_S=0)$ with n as an integer.
The relative minus sign for the interference term of $(even)^{++}$ and
$(odd)^{++}$ partial waves with M=1 and M=-1 is also due to a property of
CG coefficients.

Each partial wave amplitude $A_{J^{PC}}$ includes contributions from various
intermediate states (n), i.e.,
\be
A_{J^{PC}}=\sum_n C_nA_{J^{PC}\to n}
\ee
where $C_n$ are free complex parameters to fit the data.
In the present analysis, only $f_2(1270)\eta$,
$a_2(1320)\pi$, $a_0(980)\pi$, $\sigma\eta$, $f_0(980)\eta$ and
$f_0(1500)\eta$ intermediate states are considered.
Amplitudes $A_{J^{PC}\to n}$ are constructed from relativistic Lorentz
covariant tensors, Breit-Wigner functions and Blatt-Weisskopf barrier
factors [17].
The amplitudes used for $f_0(1500)\eta$ and $f_2(1270)\eta$ intermediate
states in our final fit are the following:
\bea
& A_{0^{-+}\to f_0\eta} &= G_{f_0}  , \\
& A_{0^{-+}\to f_2\eta} &= T^{\alpha\beta}\tilde{t}^{(2)}_{\alpha\beta}
B_2(k)G_{f_2} ,\\
& A_{2^{-+}\to f_0\eta} &= \phi^{\alpha\beta}(0)\tilde{t}^{(2)}_{\alpha\beta}
B_2(k)G_{f_0} ,\\
& A_{2^{-+}\to f_2\eta(l=0)} &= \phi^{\alpha\beta}(0)
T_{\alpha\beta}G_{f_2} ,\\
& A_{2^{-+}\to f_2\eta(l=2)} &= \phi^{\alpha\beta}(0)
\tilde{t}^{(2)}_{\alpha\gamma} T^\gamma_\beta B_2(k) G_{f_2} ,\\
& A^M_{1^{++}\to f_0\eta} &= \phi^\alpha (M)\tilde{t}^{(1)}_\alpha
B_1(k)G_{f_0}  , \\
& A^M_{1^{++}\to f_2\eta(l=1)} &= \phi_\alpha(M)\tilde{t}^{(1)}_\beta
T^{\alpha\beta} B_1(k)G_{f_2} , \\
& A^M_{1^{++}\to f_2\eta(l=3)} &= \phi^\alpha(M)\tilde{t}^{(3)}
_{\alpha\beta\gamma}
T^{\beta\gamma} B_3(k)G_{f_2} , \\
& A^M_{2^{++}\to f_2\eta(l=1)} &= \phi_{\mu\alpha}(M)
\epsilon^{\alpha\beta\gamma\delta}P_\beta\tilde{t}^{(1)}_\gamma T_\delta^\mu
B_1(k)G_{f_2} ,\\
& A^M_{2^{++}\to f_2\eta(l=3)} &= \phi_{\mu\alpha}(M)
\epsilon^{\alpha\beta\gamma\delta}P_\beta\tilde{t}^{(3)\mu\nu}_\gamma
T_\delta^\nu B_3(k)G_{f_2} ,\\
& A^M_{3^{++}\to f_0\eta} &= \phi^{\alpha\beta\gamma}(M)k_\alpha k_\beta
k_\gamma B_3(k)G_{f_0}  , \\
& A^M_{3^{++}\to f_2\eta(l=1)} &= \phi^{\alpha\beta\gamma}(M)
\tilde{t}^{(1)}_\alpha T_{\beta\gamma} B_1(k)G_{f_2}  , \\
& A^M_{3^{++}\to f_2\eta(l=3)} &= \phi^{\alpha\beta\gamma}(M)
\tilde{t}^{(3)}_{\alpha\beta\delta} T_\gamma^\delta B_3(k)G_{f_2}  , \\
& A^M_{4^{++}\to f_2\eta(l=3)} &= \phi^{\mu\nu\lambda\alpha}(M)k_\mu k_\nu
\epsilon_{\alpha\beta\gamma\delta}P^\beta k^\gamma T^\delta_\lambda
B_3(k)G_{f_2}
\eea
where $k_\mu$ is the four-momentum of the $\eta$,
$G_{f_0}=(M^2_{f_0}-s_{\pi\pi}-iM_{f_0}\Gamma_{f_0})^{-1}$ and
$G_{f_2}=(M^2_{f_2}-s_{\pi\pi}-iM_{f_2}\Gamma_{f_2})^{-1}$
are Breit-Wigner propagators for $f_0$ and $f_2$. $T_{\mu\nu}$ is a rank-2
tensor for $f_2$ and is formed by the four-momentum (p) of $f_2$ and
its break-up four-momentum (q) as
\be
T_{\mu\nu}=[ q_\mu q_\nu - {1\over 3}
(g_{\mu\nu}-{p_\mu p_\nu\over s_{\pi\pi}})q^2 ]B_2(q) .
\ee
The Blatt-Weisskopf barrier factors $B_l(k)$ with a radius of 1 fm,
the rank-$l$ tensors $\tilde{t}^{(l)}_{\delta_1\cdots\delta_l}$ for
pure $l$-wave orbital angular momentum of the $\eta f_{0,2}$ system, and the
spin-J wave functions
$\phi^{\delta_1\cdots\delta_J}(M)$ are standard as given in [17].

For $f_0(980)\eta$ and $\sigma\eta$ intermediate states, the formulae are
the same as for $f_0(1500)\eta$ except for a different $G_{f_0}$ for which
we take the parameterization of Ref. [18], i.e.,
\be
G_{f_0(980)}={1\over M_R^2-s_{\pi\pi}-ig_\pi\sqrt{1-4m_\pi^2/s_{\pi\pi}}
-ig_K\sqrt{1-4m_K^2/s_{\pi\pi}}}
\ee
with $M_R=0.99$ GeV, $g_\pi=0.117$ GeV$^2$, $g_K=0.273$ GeV$^2$,
$m_\pi=0.135$ GeV
and $m_K=0.496$ GeV;
\be
G_\sigma={1+C_0s_{\pi\pi}\over M_\sigma^2-s_{\pi\pi}
-iM_\sigma(\Gamma_1(s_{\pi\pi})+\Gamma_2(s_{\pi\pi}))},
\ee
where $C_0$ is a complex constant to be fitted by the data,
\bea
\Gamma_1(s) &=& G_1{\sqrt{1-4m_\pi^2/s}\over
\sqrt{1-4m_\pi^2/M^2_\sigma}}\cdot{(s-m^2_\pi/2)\over (M^2_\sigma -m^2_\pi/2)}
e^{-(s-M^2_\sigma)/4\beta^2}\\
\Gamma_2(s) &=& G_2{\sqrt{1-16m_\pi^2/s}\over 1+exp(\Lambda(s_0-s))}\cdot
{1+exp(\Lambda(s_0-M^2_\sigma))\over \sqrt{1-16m_\pi^2/M^2_\sigma}}
\eea
with $M_\sigma=1.067$ GeV, $G_1=1.378$ GeV, $\beta=0.7$ GeV,
$G_2=0.0036$ GeV, $\Lambda=3.5$ GeV$^{-2}$ and $s_0=2.8$ GeV$^2$.

For $a_0\pi$ and $a_2\pi$ intermediate states, the formulae are
similar to those for $f_0\eta$ and $f_2\eta$, but need symmetrization
for two pions.
The Breit-Wigner propagators for $a_0$, $a_2$, $f_0(1500)$ and $f_2$
assume constant widths. The masses and widths
(M, $\Gamma$) for $a_0$ and $f_0(1500)$ are fixed to be
($0.9834$, $0.085$) GeV and ($1.495$, $120$), respectively.
Those for $a_2$ and $f_2$ are adjusted to fit the data.
Based on these formulae, the data at each momentum are fitted by the
maximum likelihood method.

It is possible that the process $\bar pp \to \eta \pi ^0\pi ^0$ is driven,
at least partially, by $t$-channel Regge exchanges.
Even so, by Watson's theorem, each partial wave will acquire the phase
variation of any $s$-channel resonance which is present; that is, amplitudes
will contain singularities due to both $s$- and $t$-channel poles.
Our strategy will be to express $T$ matrices for individual partial waves
$T_{L,J}$ as sums over $s$-channel resonances. The formulae we use are
\bea
\sigma_{J^{PC}\to n}(s) &=& N {k_n\over s k_i} |A_{J^{PC}\to n}(s)|^2 , \\
A_{J^{PC}\to n}(s) &=& \sum_j {B_L(k_i)\Lambda_{nj}B_l(k_n)\over M_{nj}^2-s
-iM_{nj}\Gamma_{nj}} ,
\eea
where $s=M^2_{\bar pp}=M^2_{\pi\pi\eta}$, N is the normalization constant,
$k_i$ and $k_n$ are the center-of-mass momenta of initial state and channel
n respectively;
$B_L$ and $B_l$ are barrier factors for the initial state and state n
respectively;
$\Lambda_{nj}$ are complex fitting parameters; $M_{nj}$ and $\Gamma_{nj}$
are masses and widths for resonances to be fitted.
This prescription builds in the required threshold behaviour in
each partial wave. By using a sum of resonances, we satisfy the
constraint of analyticity.

\section{Results for partial waves}
The fit is shown as histograms in Figs. 5-6 for the mass spectra.
It is obviously not perfect as regards broad, slowly varying components in the
$\pi \pi $ projection of Fig. 6.
However, since we are mainly interested in scanning the
larger components from $f_2(1270)\eta$, $\sigma\eta$, $a_2\pi$ and
$a_0(980)\pi$ intermediate states, we ignore those smaller contributions
for the present study.

\begin{figure}[htbp]
\begin{center}\hspace*{-0.cm}
\epsfysize=18cm
\epsffile{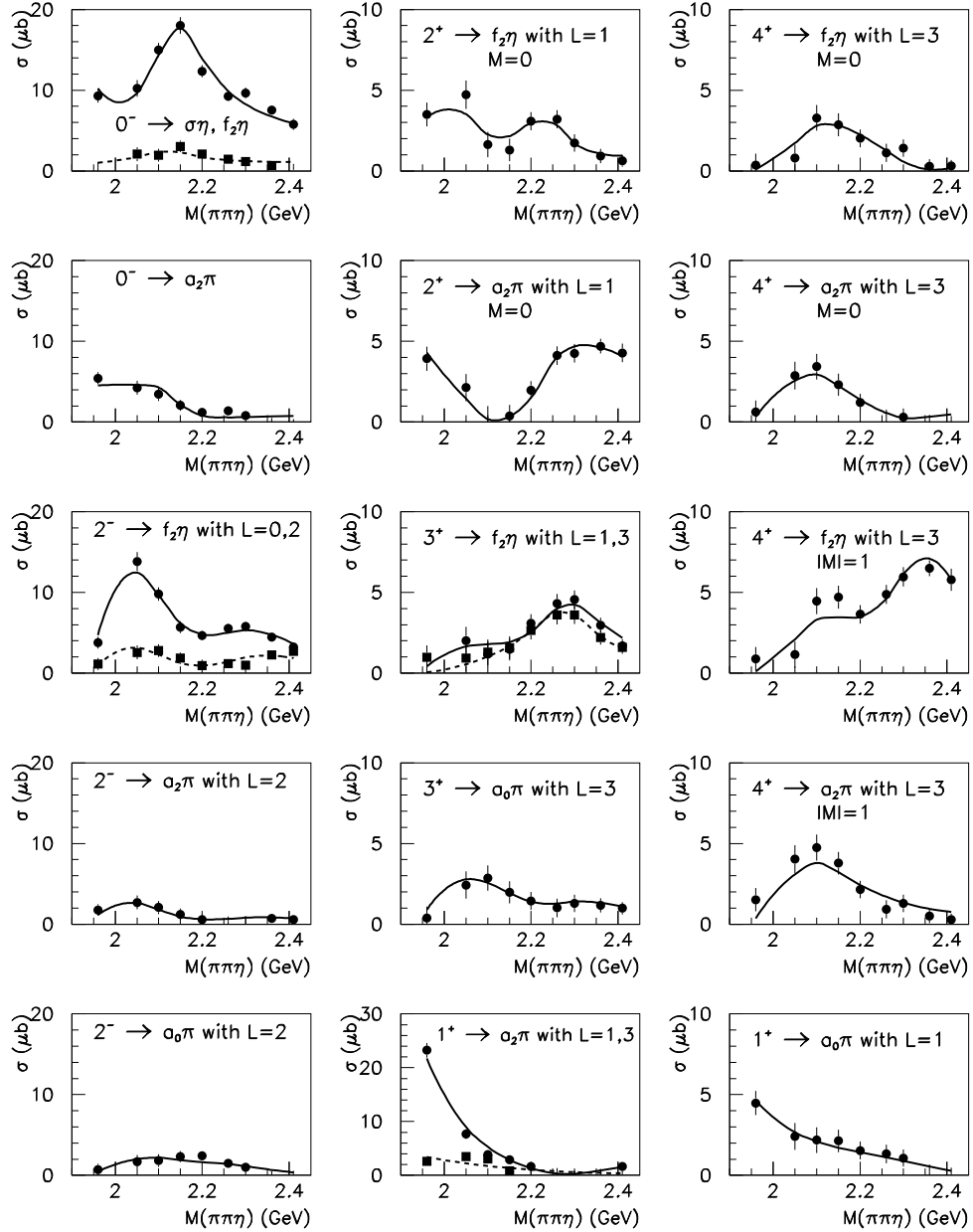}
\end{center}
\vskip -8mm
\caption{Cross sections for partial waves making the largest contributions
to $\bar pp\to\pi^0\pi^0\eta$ with $\eta\to\gamma\gamma$. For diagrams
with two components, the first label corresponds to the bigger component.
The curves are the fit to the data points in the figure and the relative
phases between components.}
\end{figure}

\begin{figure}[htbp]
\begin{center}\hspace*{-0.cm}
\epsfysize=18cm
\epsffile{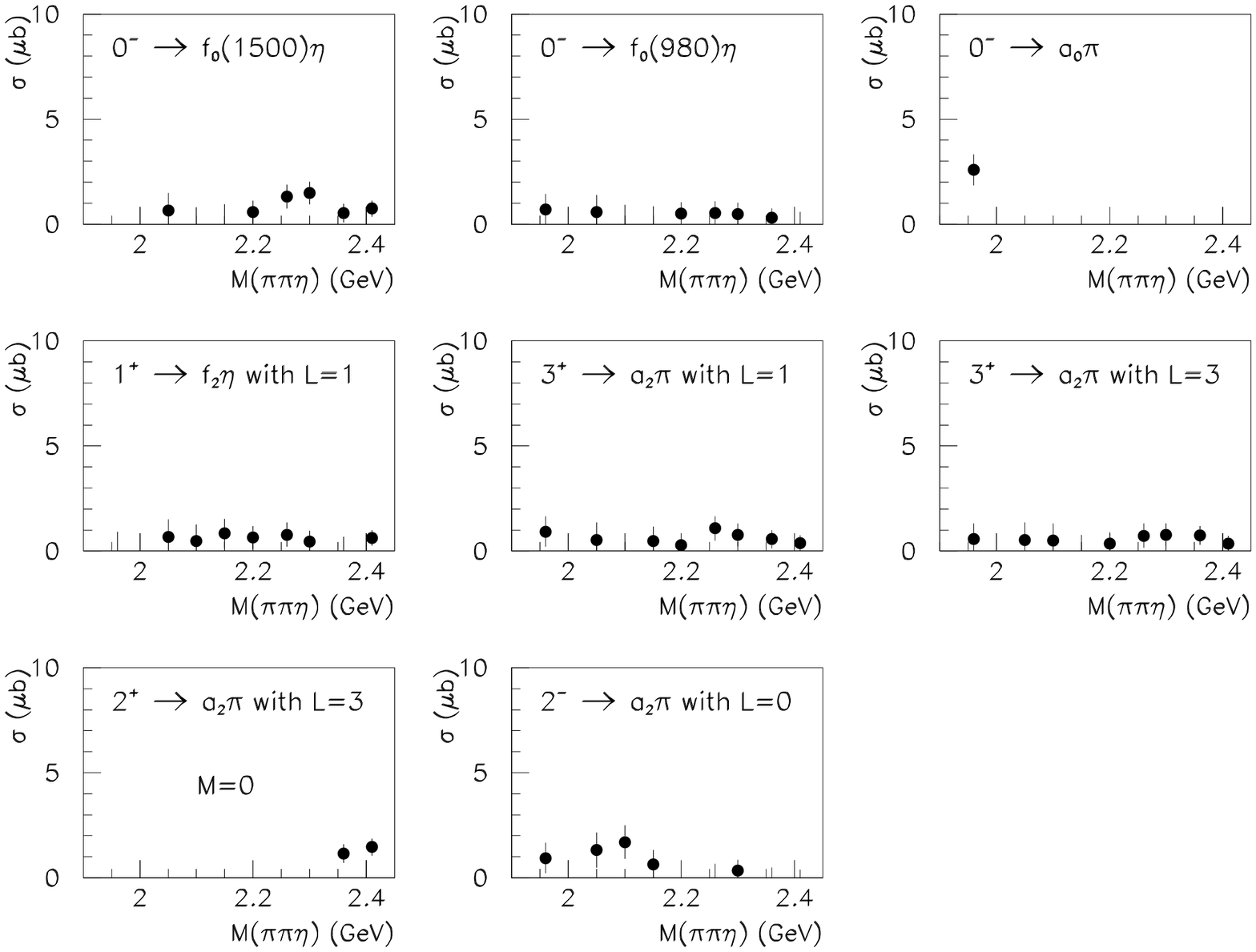}
\end{center}
\vskip -50mm
\caption{Cross sections for partial waves included in the final fit but giving
smaller contributions than those in Fig. 11 to $\bar pp \to \pi^0\pi^0\eta$ with
$\eta \to \gamma\gamma$.}
\end{figure}

The intensities of dominant partial waves are displayed in Fig. 11,
and we shall discuss a fit to them below.
The data points with error bars shown in Figs. 11 and 12 are our final fitted
results for the partial wave cross sections $\sigma_{J^{PC}\to n}$ at each
momentum for $\bar pp\to\pi^0\pi^0\eta$ with $\eta\to\gamma\gamma$.
Small waves are displayed in Fig. 12.
Partial waves with less
significant contribution than those in Fig.12 are dropped from our final fit.
Table 2 shows the masses and widths of resonances included in the fit.
Errors cover the range
of values observed in a large variety of fits. The $f_1(1700)$ is
below the range of masses accessible here, so its parameters are only
approximate.

\begin{table}[htp]
\begin{center}
\begin{tabular}{ccccccc}
\hline
$J^{PC}$ & Mass (MeV) & Width (MeV) &
${\Gamma_{\bar pp}\Gamma_{f_2\eta}\over\Gamma_{tot}^2}\cdot 10^3$ &
${\Gamma_{\bar pp}\Gamma_{a_2\pi}\over\Gamma_{tot}^2}\cdot 10^3$ &
${\Gamma_{\bar pp}\Gamma_{\sigma\eta}\over\Gamma_{tot}^2}\cdot 10^3$ &
${\Gamma_{\bar pp}\Gamma_{a_0\pi}\over\Gamma_{tot}^2}\cdot 10^3$  \\\hline
$4^{++}$ & $2044$ & $208$ & $0.54\pm 0.14$ &
$5.1\pm 0.8$ & - & - \\
$4^{++}$ & $2320\pm 30$ & $220\pm 30$ & $1.3\pm 0.4$ &
$0.6\pm 0.6$ & - & - \\
$3^{++}$ & $2000\pm 40$ & $250\pm 40$ & $0.12\pm 0.08$ &
$0.6\pm 0.6$ &  & $0.23\pm 0.11$ \\
$3^{++}$ & $2280\pm 30$ & $210\pm 30$ & $1.7\pm 0.4$ &
$4.5\pm 2.6$ &  & $0.23\pm 0.19$ \\
$2^{++}$ & $2020\pm 50$ & $200\pm 70$ & $2.1\pm 0.4$ &
$4.3\pm 1.2$ & - & -  \\
$2^{++}$ & $2240\pm 40$ & $170\pm 50$ & $2.5\pm 0.6$ &
$1.6\pm 1.6$ & - & - \\
$2^{++}$ & $2370\pm 50$ & $320\pm 50$ & $0.88\pm 0.64$ &
$16\pm 5$ & - & - \\
$1^{++}$ & $\sim 1700$ & $\sim 270$ &  &  &  & \\
$1^{++}$ & $2340\pm 40$ & $340\pm 40$ & $0.6\pm 0.6$  &
$60\pm 30$ &  & $0.84\pm 0.53$  \\
$0^{-+}$ & $2140\pm 30$ & $150\pm 30$ & $1.9\pm 1.7$ &
$6\pm 6$ & $10\pm 5$ &  \\
$2^{-+}$ & $2040\pm 40$ & $190\pm 40$ & $3.0\pm 0.3$ &
$5.0\pm 2.1$ &  & $0.4\pm 0.2$ \\
$2^{-+}$ & $2300\pm 40$ & $270\pm 40$ & $2.8\pm 0.7$ &
$2.0\pm 2.0$ &  & $0.5\pm 0.5$ \\\hline
\end {tabular}
\caption { Summary of fitted masses, widths and branching ratios corrected
for their unseen decay modes. The mass and width of $f_4(2050)$ are fixed
at
PDG values, and the status of the $0^-$ state at 2140 MeV is questionable,
as
discussed in the text. The $f_1(1700)$ is beyond the accessible mass
range.
All states have $I=0$, $G=+1$.}
\end{center}
\end{table}

The relative phases of the partial waves at each momentum are shown in
Fig.13.
Since there is no interference between spin singlet and spin triplet, or
between M=0 and M=1 for spin triplet, there will be one overall phase
undetermined for each M of spin triplet and for spin singlet.
Hence we can only determine  relative phases from our partial wave analysis.
For spin singlet ($0^-$ and $2^-$), the phases are relative to
the partial wave of $2^-\to f_2\eta$ with L=0. For spin triplet with M=0,
the phases are shown relative to $4^+\to a_2\pi$ with L=3. For spin triplet
with $|M|=1$, the phases are relative to $4^+\to f_2\eta$ with L=3.

\begin{figure}[htbp]
\begin{center}\hspace*{-0.cm}
\epsfysize=18cm
\epsffile{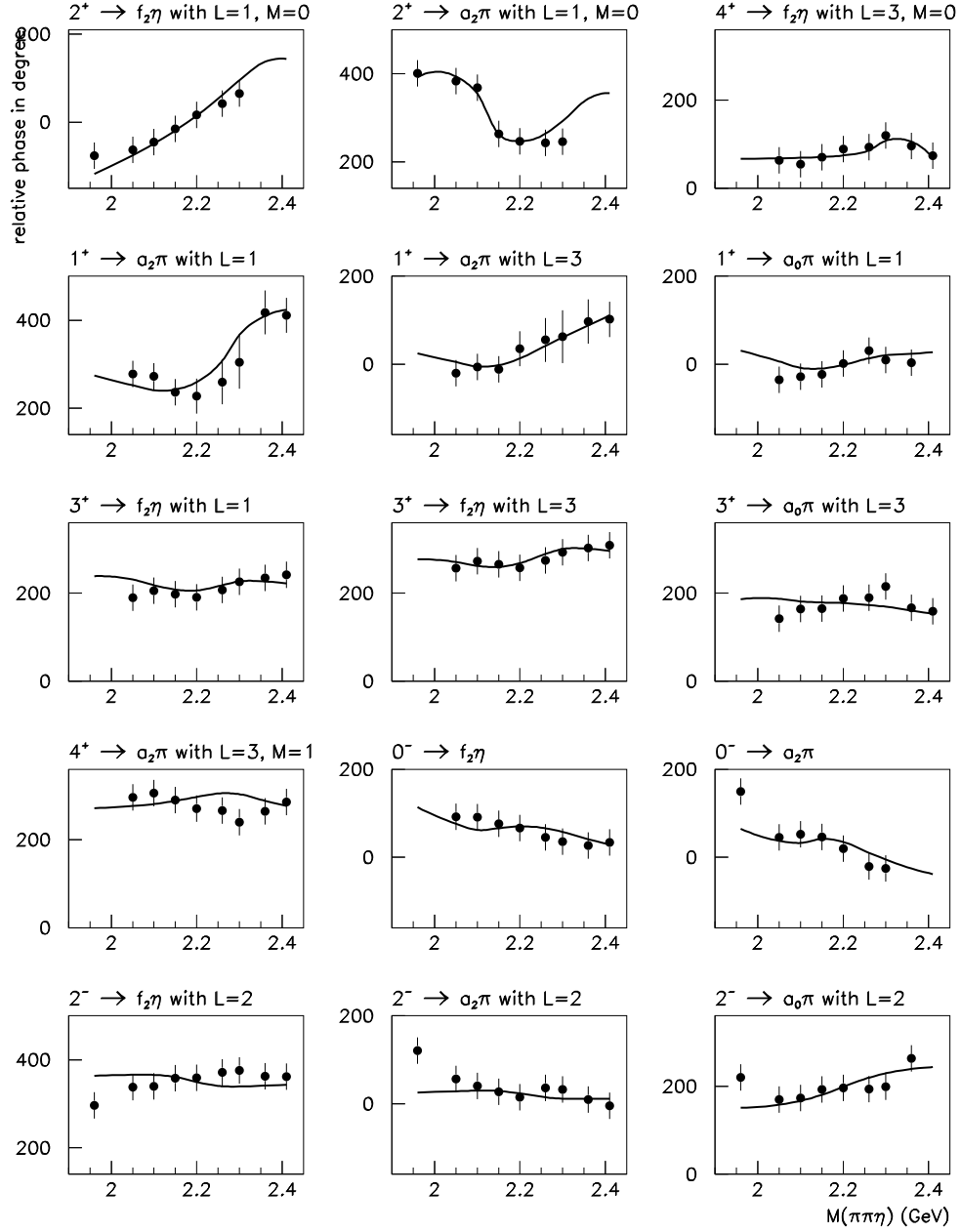}
\end{center}
\caption{Relative phases (data points with error bars) obtained from the
partial wave analysis and used for the fit (curves) to get Argand plots
together with masses and widths of the resonances. The phases for $2^+$
and $4^+$ with M=0 are relative to $4^+\to a_2\pi$ with L=3 and M=0;
the phases for $1^+$, $3^+$ and $4^+$ with M=1 are relative to
$4^+\to f_2\eta$ with L=3 and M=1; the phases for $0^-$ and $2^-$ are
relative to $2^-\to f_2\eta$ with L=0.}
\end{figure}

\subsection {$J^P = 4^+$}
For $4^{++}$, a peak around 2090 MeV is clear for all $4^{++}$ channels.
It can be fitted  by a Breit-Wigner amplitude with the mass
and width fixed to the PDG values for the well established $4^+$ resonance
$f_4(2050)$. The shift of the peak position to 2090 MeV is due
to the centrifugal barrier factors for both initial and final states.
Its decays into $f_2\eta$ and $a_2\pi$ appear with comparable strength in the
$\eta \pi ^0 \pi ^0$ channel.

In addition to the $f_4(2050)$, there is clearly another $4^{++}$ peak
around $2.32$ GeV in $4^+\to f_2\eta$ in the M=1 partial wave.
This resonance may be identified with $f_4(2300)$ of the PDG, observed earlier
in many analyses of $\bar pp \to \pi ^- \pi ^+$.
The mass, width and phase with respect to $f_4(2050)$ are adjusted freely.
The mass optimises at $M = 2320 \pm 30 $ MeV and the width at $\Gamma =
220 \pm 30$ MeV.
These agree closely with earlier determination quoted by the PDG,
and also with
recent VES data on  $\eta\pi^+\pi^-$ in the $\pi A$ reaction [19].
The latter find $M = 2330 \pm 10(stat) \pm 20(syst)$ MeV, $\Gamma = 225
\pm 20 \pm 40$ MeV.
They also observe this resonance in $\omega \omega$ data [20].
The $f_4(2300)$
is also observed in our data on $\bar pp \to \pi ^0 \pi ^0$ [15], with
a slightly lower mass of 2295 MeV.
The $f_4(2300)$ resonance acts as a valuable interferometer, determining
the phases
of $3^+$, $2^+$ and $1^+$ amplitudes over the mass range 2150--2400 MeV.

\begin{figure}[htbp]
\begin{center}\hspace*{-0.cm}
\epsfysize=9cm
\epsffile{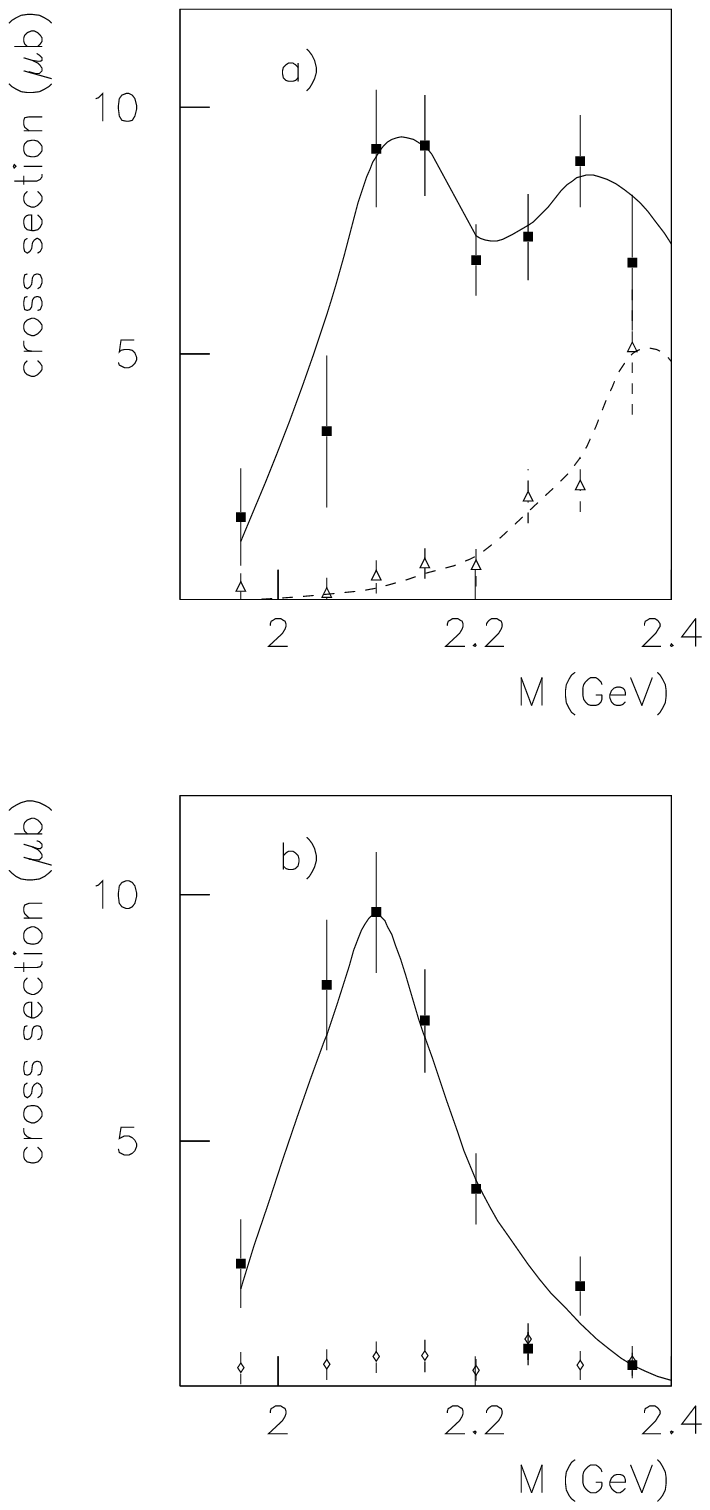}
\vskip -90.41mm
\hskip 1.35mm
\epsfysize=9cm
\epsffile{3F43H4.PS}
\end{center}
\vskip -8mm
\caption{Contributions to (a) $f_2(1270)\eta$, (b) $a_2(1320)\pi$ from
$^3F_4$ (black squares) and $^3H_4$ (open triangles).}
\end{figure}

From the $M = 1$ and $M = 0$ amplitudes for $4^+$, we reconstruct the linear
combinations for $^3F_4$ and $^3H_4$.
Their intensities are shown in Fig. 14 for $f_2(1270)\eta$ and $a_2(1320)\pi$
channels.
The $f_4(2050)$ resonance is almost purely $^3F_4$.
The $a_2\pi$ channel is fed mostly by $f_4(2050)$ with a possible
weak contribution from $f_4(2320)$;
the $^3H_4$ contribution to $a_2\pi$ is barely significant.
In contrast, the $f_2\eta$ channel is fed by both $f_4(2050)$ and
$f_4(2320)$ and the latter has a strong $^3H_4$ component. This is
in agreement with the analysis of $\bar pp \to \pi ^- \pi ^+$
by Hasan and Bugg [6]; their Fig. 3 shows a strong $^3H_4$ component in
$f_4(2320)$.
The VES collaboration [19]  finds that
$f_4(2320)$ decays dominantly to $f_2\eta$, in agreement with present results.

\subsection {$J^P = 3^+$}
For $J^{PC} = 3^{++}$, there are significant enhancements at low mass $(M
\simeq 2000$ MeV) in both $a_0(980)\pi$ and $f_2(1270)\eta$ with $L = 1$.
At high mass ($M \simeq 2280$ MeV) there is a strong peak in $f_2(1270)\eta$
decays with both $L = 1$ and $L = 3$ decays.
Fitted masses and widths are given in Table 2.
There are no earlier listings of these resonances by the PDG.
The observed phase with respect to $f_4(2050)$ and $f_4(2300)$ shown in Fig. 13
obviously requires the presence of at least one $3^+$ resonance, and is
poorly fitted without two.
The Argand diagram is shown in fig. 15.

\begin{figure}[htbp]
\begin{center}\hspace*{-0.cm}
\epsfysize=18cm
\epsffile{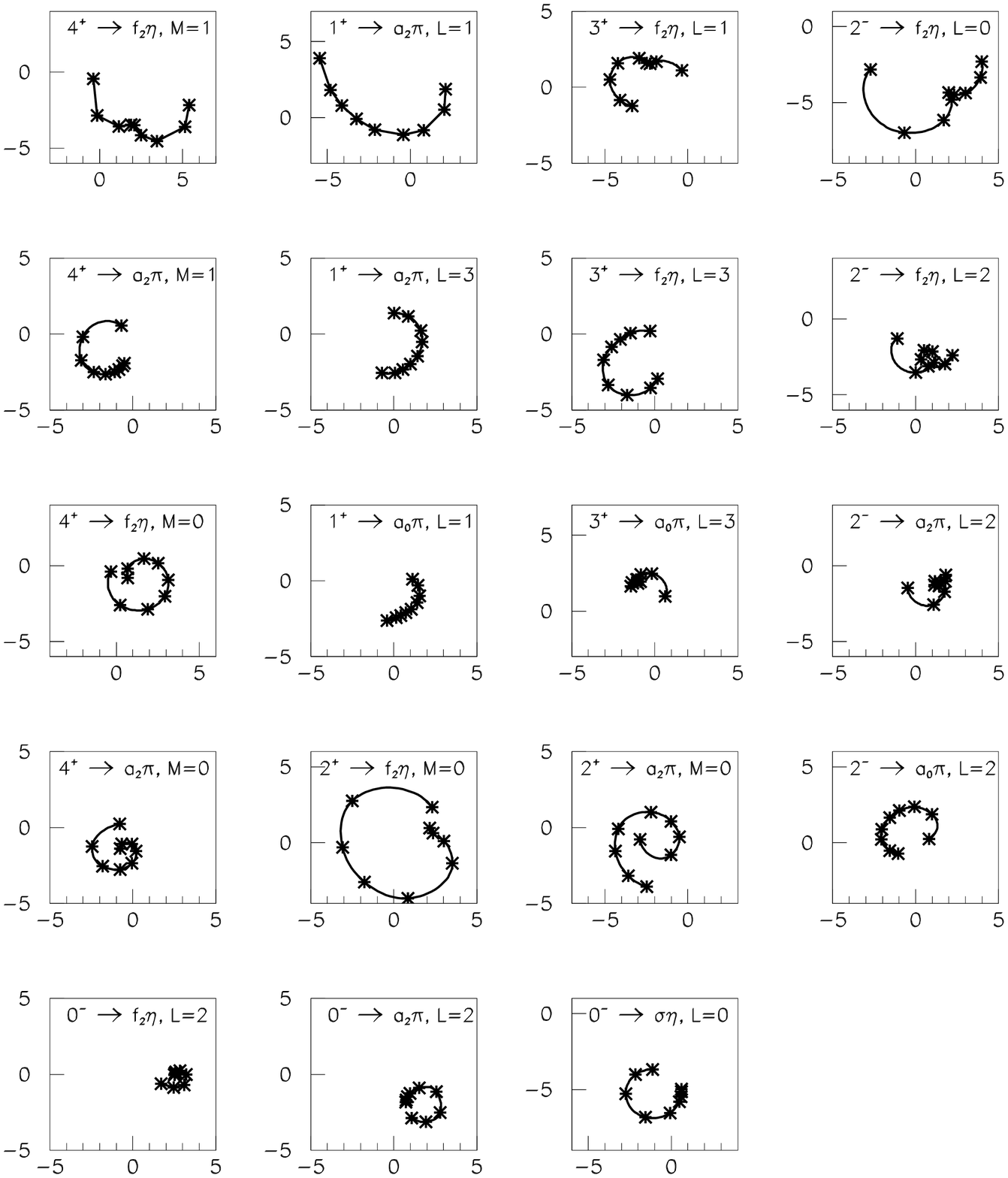}
\end{center}
\vskip -8mm
\caption{Argand plots corresponding to curves in Figs.11 and 13.}
\end{figure}

\subsection {$J^P = 2^+$}
For $2^{++}$, there is a peak in $f_2(1270)\eta$ at $\sim 2020$ MeV and a
peak at low masses in $a_2(1320)\pi$.
At high mass around 2300 MeV, there is a strong peak in the $a_2(1320)\pi$
channel.
In $f_2(1270)\eta$, there is a further peak at $\sim 2230$ MeV.
The obvious question is how many resonances are required to fit these diverse
structures.
The phase variation observed on the Argand diagram, Fig. 15, requires at least
two resonances from the observed 360$^{\circ}$ phase advance.

We find that the fit is poor without three resonances.
The lowest peak fits naturally to a resonance with $M = 2020  \pm 50$ MeV,
$\Gamma = 200 \pm 70$ MeV.
Our data on $\bar pp \to \pi ^0 \pi ^0$ independently find a resonance
at 2020 MeV [15],
and the analysis of Hasan and Bugg [6] of data on $\bar pp \to \pi ^- \pi ^+$
likewise finds an $f_2$ resonance at 1996 MeV.
We have tried an alternative fit using instead $f_2(1920)$ observed by both
GAMS [21] and VES [22] collaborations.
The $2^{++}\to a_2\pi$ partial wave can be reproduced
equally well with this assignment,
but the $2^{++}\to f_2\eta$ partial wave is seriously underfitted
by a factor 3 at 2050 MeV, ruling out a fit by $f_2(1920)$ only.

The proximity of this resonance to $f_4(2050)$ suggests that it may be
identified as the $\bar qq$ $^3F_2$ state expected near this mass.
Because $2^+$ amplitudes with $M = 1$ are negligible,
$^3P_2$ and $^3F_2$ amplitudes have the same $s$-dependence; $^3F_2$ is the
larger by a factor 1.44.
This strong coupling of $f_2(2030)$ to $\bar pp$ $^3F_2$ also suggests
identification with $\bar qq$ $^3F_2$:
high $L$ in $\bar qq$ is likely to be associated with high $L $ in decay
channels, because of the peaking of wave functions at large $r$.

At higher masses, a fit with a single resonance, shown by the dashed curve
in Fig. 16, is much poorer than with two separate resonances.
The peak at 2240 MeV in $f_2(1270)\eta$ has a mass compatible with $\xi(2230)$
observed in $J/\Psi$ radiative decays
[23], but has a larger width
of about 170 MeV.
This resonance may be interpreted as the $n = 4$ $\bar qq$
$^3P_2$ state, in the sequence $f_2(1270)$, $f_2(1565)$, $f_2(1920)$,
$f_2(2240)$.
The $f_2(2370)$ finds a natural explanation as $n = 2$ $\bar qq$ $^3F_2$, i.e.
the radial excitation of $f_2(2020)$.
Its strong $L = 3$ decay supports this interpretation.
In present data, both $f_2(2240)$ and $f_2(2370)$ appear in both $^3P_2$ and
$^3F_2$, suggesting mixing between these states.


\begin{figure}[htbp]
\begin{center}\hspace*{-0.cm}
\epsfysize=8cm
\epsffile{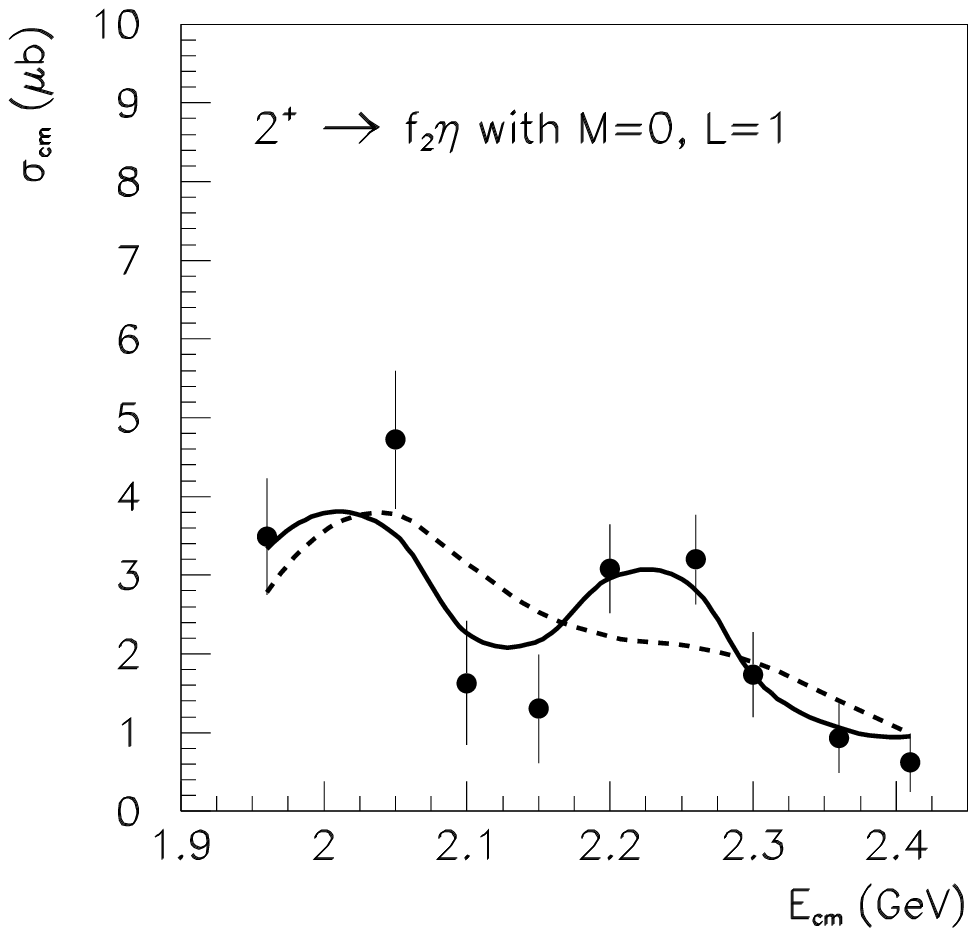}
\vskip -8mm
\end{center}
\caption{Fits with $f_2(2240)$ (full curve) and without (dashed).}
\end{figure}

The two peaks around 2020 MeV and 2370  MeV have masses and widths
compatible with $f_2(2010)$ and $f_2(2340)$ listed by the PDG [4].
However, those observations were in the $\phi \phi$
channel and could be different resonances, e.g. $\bar ss$.
We also remark that the peak in the $\phi \phi$ data of Etkin et al. [24]
actually appears at $\sim 2150$ MeV. It is the rapid opening of the $\phi \phi
$ phase space which leads to a pole at much lower mass, 2020 MeV, in the
K-matrix fit to their data.

\subsection {$J^P = 1^+$}
For $J^{PC}= 1^{++} \to a_2(1320)\pi$ and $a_0(980)\pi$, there is a peak at
the lowest masses. This suggests a resonance  close to or
below the $\bar pp$ threshold.
However,  as discussed below, the phase variation of the $1^+$ amplitude
provides evidence for a resonance around 2340 MeV.
The phase variation shown in Fig. 15 obviously requires resonant activity in
the mass range 2000--2400 MeV.

\subsection {$J^P = 2^-$}
Partial waves with quantum numbers $2^{-+}$ and $0^{-+}$ correspond to $\bar
pp$ singlet states, and therefore there is no interference with even parity
(triplet) partial waves.
For $2^{-+}$ there is a strong peak in $f_2(1270)\eta$ at $\sim 2050$ MeV
and a smaller peak in $a_2(1320)\pi$ at similar mass.
There is evidence for a further peak at $\sim 2300$ MeV.
The lower peak is well fitted by a resonance with $M = 2040 \pm 40$ MeV.
The almost 360$^{\circ}$ phase advance observed on the Argand diagram points
strongly towards the presence of two resonances, the second at $2300 \pm 40$
MeV.
The PDG does not list any $I = 0$ $J^{PC} = 2^{-+}$ resonance in this mass
range.
A possible $I = 1$ partner is listed in the form of $\pi _2(2100)$.

\subsection {$J^P = 0^-$}
For $0^{-+}$, there is a broad, slowly varying intensity with evidence for a
strong peak superimposed at $M \sim 2140$ MeV. The slowly varying component
may correspond to the broad $0^{-+}$ object used in
describing $J/\Psi$ radiative decays to $\rho\rho$, $\omega\omega$,
$K^*\bar K^*$, $\phi\phi$ and $\eta\pi\pi$ [25].
The peak at 2140 MeV may correspond to a narrow resonance.
However, it is observed in the $\eta \sigma$ channel, which contributes across
the entire Dalitz plot.
This contribution might absorb weak components not presently fitted to the
data, for example due to $a_0(1450)$, $a_2(1660)$, $\hat \rho (1405)$ or
further resonances in the production process around 2 GeV. In view of this
possibility, the interpretation in terms of a resonance is ambiguous.
Unfortunately, the relative phase with respect to $2^-$ is not well determined,
so the phase variation cannot be used for independent evidence of resonant
activity.

\section{Final fit to the partial waves}
To get more precise values for masses and widths for resonances,
we use interfering sums of the Breit-Wigner amplitudes to fit the partial
wave cross sections in Fig.11 and the relative phases of the
partial waves in Fig.13 simultaneously.  The fit is shown in
Figs. 11-13 as full curves.

Besides the obvious resonances mentioned
in the previous section, we need another $1^{++}$ resonance at about 2340 MeV
with width $\sim 340$ MeV. Without it, we cannot describe the relative
phase
between $1^{++}$ and $4^{++}$ partial waves; also we would need the lower
$1^{++}$ resonance to be very narrow ($<50$ MeV) in order to explain the
sharply decreasing $1^{++}$ partial wave cross section. In our present fit
with two $1^{++}$ resonances, the $f_1(2340)$ amplitude
interferes destructively with the tail of the lower $1^{++}$ resonance
and causes the sharply decreasing cross section with a broad dip around
2340 MeV. The phase motion caused by this $f_1(2340)$ can be seen clearly
in the Argand plots for $1^{++}$ partial waves of Fig.15.

In Table 2, the branching ratios are calculated at the resonance masses
and are corrected for their unseen decay modes, except for $a_0(980)$
where $\Gamma_{a_0\pi}=\Gamma_{a_0\pi\to\eta\pi\pi}$.

For an ordinary $q\bar q$ state, the relative ratio $f_2\eta$/$a_2\pi$
is expected to be smaller than 0.64. This allows for the  36\% component of
$\bar ss$ in the $\eta$. The centrifugal barrier and phase space will further
suppress $f_2\eta$.
Most of the branching ratios in Table 2 are in qualitative agreement with what
is expected for $\bar qq$ states.
However, the $f_2(2230)$ has an anomalously strong branching ratio to
$f_2(1270)\eta$ compared with $a_2(1320)\pi$.

For the well-established $f_4(2050)$, only 44\% of its branching ratios are
listed in the Particle Data Tables [4], in which $\pi\pi$ has a branching
ratio of ($17\pm 1.5$)\%. In a very recent analysis [26] of $\bar
pp\to\pi\pi$,
the ratio ${\Gamma_{\bar pp}\Gamma_{\pi\pi}\over\Gamma^2_{tot}}$ was reported
to be $(2.2\sim 2.4)\times 10^{-3}$. Using this information,
we can get the branching ratios of $f_4(2050)$ to $\bar pp$, $a_2\pi$ and
$f_2\eta$ to be $(1.4\pm 0.1)\%$, $(30\pm 5)\%$ and $(3.9\pm 1.0)\%$,
respectively.

\subsection {Comments on the resonance spectrum}
The $f_4(2044)$, $f_3(2000)$, $f_2(2020)$ and $\eta_2(2040)$ cluster
closely
into a tower of resonances, as anticipated in the Veneziano model.
Likewise the $f_4(2320)$, $f_3(2280)$, $f_2(2370)$, $f_1(2340)$ and $\eta
_2(2300)$ show indications of clustering into a tower at the higher masses.

The $f_2(1920)$ originally discovered by both GAMS and VES has recently been
confirmed in further VES data with increased statistics, decaying to $\omega
\omega$ [20].
There is also a strong $f_2(1270)\eta$ signal in VES
$\eta \pi ^+ \pi ^-$ data.
Together with the $f_2(2020)$ we observe here, $f_2(2240)$ and $f_2(2370)$,
this tentatively completes the identification of the $\bar qq$ $I=0$ $^3P_2$ and
$^3F_2$ states expected in this mass range.

We conclude with some speculative suggestions of a scheme which concerns
mixing of $\bar qq$ states with the $2^+$ glueball expected in this mass range.
In our data on $\bar pp \to \eta \eta \pi ^0$ [2], there is evidence for a
further broad $f_2(1980)$ decaying to $\eta \eta$, with mass $M = 1980 \pm 50$
MeV, $\Gamma = 500 \pm 100$ MeV.
Its effects are seen clearly down to masses of $\sim 1550$ MeV.
There is also evidence for a broad $2^+$ resonance in $4\pi$ final states in
central production [27].
Such a broad state was predicted by Bugg and Zou [25].
It may be interpreted as a mixed state formed from the $2^+$ glueball, expected
at $\sim 2-2.2$ GeV, and nearby $\bar qq$ states.
Anisovich et al. [28] have argued that this mixing will lead to a broad state,
accumulating the widths of nearby $\bar qq$ states and making them narrower.
The $f_2(1920)$ and $f_2(2240)$ are indeed somwhat narrower this is usual for
resonances in this mass range.
Mixing with a glueball provides a natural explanation of the anomalous decays
of $f_2(2020)$ and $f_2(2340)$ to $\phi \phi$, observed by Etkin et al. [24].

The glueball may be small, with radius $\sim 0.3$ fm; there are
indications for this small radius in QCD Lattice calculations [29].
The small radius allows much of the glueball mass to be attributed to
zero-point energy.
Such a small object will mix preferentially with $\bar qq$ $^3P_2$ states
rather than $\bar qq$ $^3F_2$, whose wave functions are strongly localised at
large $r$.
The preferential decays of $f_2(1920)$ and $f_2(2240)$ to $f_2(1270)\eta$,
despite its smaller phase space than $a_2(1320)\pi$, may be a further indication
of mixing with the $2^+$ glueball.

\section{Summary}
In summary, we have observed a new decay mode $\eta\pi\pi$ for
$f_4(2050)$.
In addition, we have evidence for 7 new or poorly established resonances
in the energy range from 1.96 to 2.41 GeV, i.e., $f_4(2320)$, $f_3(2000)$,
$f_3(2280)$, $f_2(2240)$, $f_1(2340)$, $\eta_2(2040)$
and $\eta_2(2300)$.
They appear to cluster into two towers of resonances around 2000--2050 MeV and
2300 MeV. Results are broadly consistent with earlier evidence for
$f_4(2300)$, $f_2(2020)$ and $f_2(2340)$.

\section{Acknowledgement}
We thank the Crystal Barrel Collaboration for allowing use of the data.
We also thank the technical staff of the LEAR machine group
and of all the participating institutions for their invaluable
contributions to the success of the experiment. We acknowledge financial
support from the
the British Particle Physics and
Astronomy Research Council (PPARC).
The St.Petersburg group thanks INTAS
for financial support, contract RFBR 95-0267, and also PPARC for
financial assistance for collaborative work.


\begin{thebibliography}{99}
\bibitem{1} A. Anisovich et al., {\it Study of $\bar pp \to \pi ^0 \pi ^0 \eta$
from 600 to 1940 MeV/c}.
\bibitem{2} A. Anisovich et al., {\it Study of the process $\bar pp \to
\eta \eta \pi ^0 $ from 1350 to 1940 MeV/c}, Phys. Lett. B449 (1999) 145.
\bibitem{3} A. Anisovich et al., {\it Observation of $f_0(1770) \to \eta \eta$
in $\bar pp \to \eta \eta \pi ^0$ reactions from 600 to 1940 MeV/c}, 
Phys. Lett. B 449 (1999) 154.
\bibitem{4} Particle Data Group, C. Caso et al., Euro. Phys. J. C3 (1998) 1.
\bibitem{5} A.Hasan et al., Nucl. Phys. B378 (1992) 3.
\bibitem{6} A.Hasan and D.V.Bugg, Phys. Lett. B334 (1994) 215.
\bibitem{7} G. Veneziano, Nu. Cim. 57A (1968) 190.
\bibitem{8} G.Bali et al. (UKQCD), Phys. Lett. {\bf B307} (1993) 378;
H.Chen, J.Sexton, A.Vaccarino and D.Weingarten, Nucl. Phys. B (Proc. Suppl.)
34 (1994) 357.
\bibitem{9} V.A.Novikov, M.A.Shifman, A.I.Vainshtein and V.I.Zakhnov,
Nucl. Phys. B 191 (1981) 301.
\bibitem{10} J.Y.Cui, J.M.Wu and H.Y.Jin, Phys. Lett. B424 (1998) 381.
\bibitem{11} V.V. Anisovich et al., Phys. Lett. B323 (1994) 233; C. Amlser et
al., Phys. Lett. B355 (995) 425.
\bibitem{12} D.V.Bugg et al., Phys. Lett. {\bf B353} (1995) 378.
\bibitem{13} E. Aker et al., Nucl. Instr. A321 (1992) 69.
\bibitem{14} C.A. Baker, N.P Hessey, C.N. Pinder and C.J. Batty,
Nucl. Instr. and Methods in Phys. Res. A394 (1997) 180.
\bibitem{15} A. Anisovich et al., {\it $\bar pp \to \pi ^0 \pi ^0 $
from 600 to 1940 MeV/c}, submitted to Phys. Lett. B.
\bibitem{16} A.Bertin et al., Phys. Rev. D57 (1998) 55.
\bibitem{17} S.U.Chung, Phys. Rev. D48 (1993) 1225; D57 (1998) 431.
\bibitem{18}  D.V.Bugg, A.V.Sarantsev and B.S.Zou, Nucl. Phys. {B471}
(1996) 59.
\bibitem{19} D. Ryabchikov, AIP Conf. Proc. 432, eds. S.-U. Chung and
H.J. Willutzki, (Amer. Inst. of Phys. New York, 1998), p603.
\bibitem{20} D. Ryabchikov, private communication.
\bibitem{21} D. Alde et al., Phys. Lett. B276 (1992) 375.
\bibitem{22}  S.I.Beladidze et al., Z.Phys. C54 (1992) 367.
\bibitem{23} BES collaboration, J.Z.Bai et al.,
		Phys. Rev. Lett. {76} (1996) 3502.
\bibitem{24} A. Etkin et al., Phys. Lett. B201(1988) 568.
\bibitem{25} D.V.Bugg and B.S.Zou, Phys. Lett. B396 (1997) 295.
\bibitem{26} B.R.Martin and G.C.Oades, Preprint IFA-SP-98-1, HEPPH-9802261.
\bibitem{27} D. Barberis et al., Phys. Lett. B413 (1997) 217.
\bibitem{28} V.V. Anisovich, D.V. Bugg and A.V. Sarantsev, hep-ph/9711478
(1997) and Phys. Lett B (to be published).
\bibitem{30} D. Weingarten, private communication.
\end{thebibliography}
\end{document}